\documentclass[acmlarge]{acmart}
\usepackage{mdframed}  
\usepackage{xspace}
\usepackage{multirow}
\usepackage{enumitem}
\usepackage{soul}
\usepackage{tabularx}
\usepackage{subfigure}
\usepackage{graphicx}
\usepackage{amsmath}
\usepackage{booktabs} 
\usepackage{adjustbox}
\usepackage{rotating}
\usepackage{xcolor}

\usepackage{soul}         
\makeatletter             
\soulregister\cite7        
\soulregister\ref7
\soulregister\pageref7
\makeatother
\usepackage{hyperref}  
 
\usepackage{tabularx}
\usepackage[utf8]{inputenc}
\usepackage[most]{tcolorbox}
\usepackage{needspace}
\usepackage{etoolbox}

\makeatletter
\pretocmd{\@bibitem}{\needspace{2\baselineskip}}{}{}
\pretocmd{\@lbibitem}{\needspace{2\baselineskip}}{}{}
\makeatother

\setcopyright{cc}
\setcctype{by}
\acmJournal{IMWUT}
\acmYear{2026} \acmVolume{10} \acmNumber{1} \acmArticle{18}
\acmMonth{3} \acmPrice{} \acmDOI{10.1145/3789689}

\begin{document}
\title{AEDHunter: Investigating AED Retrieval in the Real World via Gamified Mobile Interaction and Sensing}
\author{Helinyi Peng}
\authornote{Corresponding author.}

\affiliation{%
 \institution{The University of Tokyo}
  \city{Tokyo}
 \country{Japan}}\email{penghly@outlook.com}
 \orcid{0009-0002-1521-3407}
\author{Akihito Taya}
\affiliation{%
 \institution{The University of Tokyo}
 \city{Tokyo}
 \country{Japan}}\email{taya-a@iis.u-tokyo.ac.jp}
\orcid{0000-0001-9074-9709}

\author{Yuuki Nishiyama}
\affiliation{%
 \institution{The University of Tokyo}
 \city{Tokyo}
 \country{Japan}}\email{nishiyama@csis.u-tokyo.ac.jp}
 \orcid{0000-0002-5549-5595}
 
\author{Kaoru Sezaki}
\affiliation{%
 \institution{The University of Tokyo}
 \city{Tokyo}
 \country{Japan}}\email{sezaki@iis.u-tokyo.ac.jp}
 \orcid{0000-0003-1194-4632}
\renewcommand{\shortauthors}{Peng et al.}

\authorsaddresses{%
Authors’ Contact Information: {\href{https://orcid.org/0009-0002-1521-3407}{Helinyi Peng}} (corresponding author), The University of Tokyo, Tokyo, Japan, penghly@outlook.com; {\href{https://orcid.org/0000-0001-9074-9709}{Akihito Taya}}, The University of
Tokyo, Tokyo, Japan, taya-a@iis.u-tokyo.ac.jp; {\href{https://orcid.org/0000-0002-5549-5595}{Yuuki Nishiyama}},The University of Tokyo, Tokyo, Japan, nishiyama@csis.u-tokyo.ac.jp;
{\href{https://orcid.org/0000-0003-1194-4632}{Kaoru Sezaki}}, The University of Tokyo, Tokyo, Japan, sezaki@iis.u-tokyo.ac.jp.
}

\begin{abstract}
Early defibrillation significantly improves survival rates in cases of out-of-hospital cardiac arrest. However, limited public awareness of Automated External Defibrillator (AED) locations constrains their effective use. Existing solutions, such as static 2D maps, often fall short in urgent or complex real-world scenarios. To address this challenge, we developed AEDHunter, a gamified, location-based mobile application designed to transform AED retrieval into an engaging and repeatable practice experience. Leveraging smartphone sensors to analyze participants’ movement and learning patterns, and using low-cost Bluetooth tags to verify arrivals at AED locations, AEDHunter guides users through multiple sessions of AED discovery. In a real-world evaluation study, participants significantly reduced their AED retrieval times after repeated practice sessions and reported increased confidence in locating AEDs. Additionally, we employ a two-state activity detector to identify ``exploratory pauses'', which are then used as a behavioral learning signal to quantify hesitation and its progressive reduction through practice. 
Our findings suggest that gamified applications like AEDHunter can improve AED retrieval performance through repeated, in-situ training and enhance self-reported preparedness, offering design insights for technology-supported learning and public safety applications.
\end{abstract}


\maketitle
\section{INTRODUCTION}

Out-of-hospital cardiac arrest (OHCA) refers to the sudden cessation of cardiac mechanical activity occurring outside a hospital setting, leading to an immediate loss of consciousness and the absence of both pulse and respiratory function~\cite{roger2011heart,welbourn2018does}. 
It remains one of the leading causes of mortality worldwide, with survival rates persistently low~\cite{kiguchi2020out}. Early defibrillation is crucial, as each minute of delay decreases the likelihood of survival by 7--10\%~\cite{ong2014geographic,larsen1993predicting}. Despite extensive efforts to increase the availability of Automated External Defibrillators (AEDs) in public spaces, their use during emergencies remains alarmingly limited. For example, in South Korea and Germany, AEDs are employed in only 0.7\% and 1.4\% of OHCA cases, respectively~\cite{pommerenke2023automated,lee2021public}. This underutilization represents a critical missed opportunity for life-saving interventions.

Even in well-marked locations, low AED utilization rates suggest that the issue extends beyond visibility and includes limited public awareness as well as unfamiliarity with how to access AEDs~\cite{zinckernagel2016qualitative,smith2017barriers}. Surveys indicate that only 5--22\% of non-professional individuals can locate their nearest public-access AED~\cite{brooks2015public, fan2016public, kozlowski2013knowledge}, and alarmingly, only 32\% of sixth-year medical students know the location of the nearest AED to their home~\cite{timler2024automatic}. These findings highlight the first challenge (C1\label{challenge-1}) in AED retrieval: individuals often overlook AEDs embedded in their everyday environments, leading to widespread unawareness of their locations. Therefore, interventions are needed to increase the public salience of AEDs and promote active engagement with their locations. Enhancing bystander familiarity through repeated exposure and interaction is essential for improving preparedness and enabling timely AED retrieval when needed.

Although several existing websites and applications provide data on AED locations, merely accessing this information is often insufficient for effectively managing real emergency situations. Most AEDs are installed indoors, where navigation can be challenging due to weak GPS signals, complex building layouts, and limited spatial orientation cues~\cite{neves2019automated, bin2019mobile}. Under high-stress conditions, interpreting two-dimensional maps becomes cognitively demanding, which can significantly reduce the likelihood of timely AED retrieval. This highlights a second critical challenge (C2\label{challenge-2}): the inherent difficulty of indoor navigation and spatial orientation during emergencies. Unlike outdoor scenarios, indoor navigation cannot rely solely on conventional mapping solutions. To mitigate this limitation, systems should go beyond simply providing AED location data and instead incorporate active interventions that enhance users’ spatial familiarity and cognitive readiness. For example, gamified smartphone applications or interactive training tools could help individuals develop spatial memory and familiarity with indoor AED locations, thereby reducing cognitive load and improving retrieval speed in time-critical situations. Repeated, low-stakes in-situ practice can build spatial familiarity and procedural memory that may transfer to time-critical retrieval under stress.	

Moreover, few studies have empirically examined whether interventions that increase awareness of nearby AEDs improve in-situ AED retrieval(C3\label{challenge-3}).
Although some gamified approaches encourage people to visit AED locations~\cite{merchant2013crowdsourcing, hao2019fun, bin2019mobile}, clear evidence from real-world evaluations with objective measures remains limited. For example, our prior work introduced a gamified mobile application that encourages users to visit and activate AEDs~\cite{peng2025aedhunter}, but the initial evaluation primarily relied on subjective self-reports rather than objective, in-situ performance measures. Understanding the actual impact of these interventions is critical for guiding effective AED awareness strategies and ultimately enhancing bystander responses.

Lastly, there are limited insights regarding fine-grained, real-world AED retrieval behaviors (C4). Prior studies have described AED retrieval as a sequence of distinct phases or interconnected events~\cite{jonsson2020brisk}. These steps include initial preparation, such as reading maps and planning routes, followed by locating buildings equipped with AEDs, and finally searching for AEDs within indoor environments~\cite{peng2024poster}. 
However, while this phase structure has been articulated, objective, sensor-based phase-level measurement in real-world retrieval remains limited. As highlighted by Dorian et al.~\cite{dorian2020retrieving}, understanding all the steps involved in finding an AED and the time required is crucial for identifying effective strategies to overcome barriers to successful retrieval. This gap limits the development of effective interventions and the improvement of AED utilization.

\begin{figure}[tb]
  \centering
  \includegraphics[width=1\linewidth]{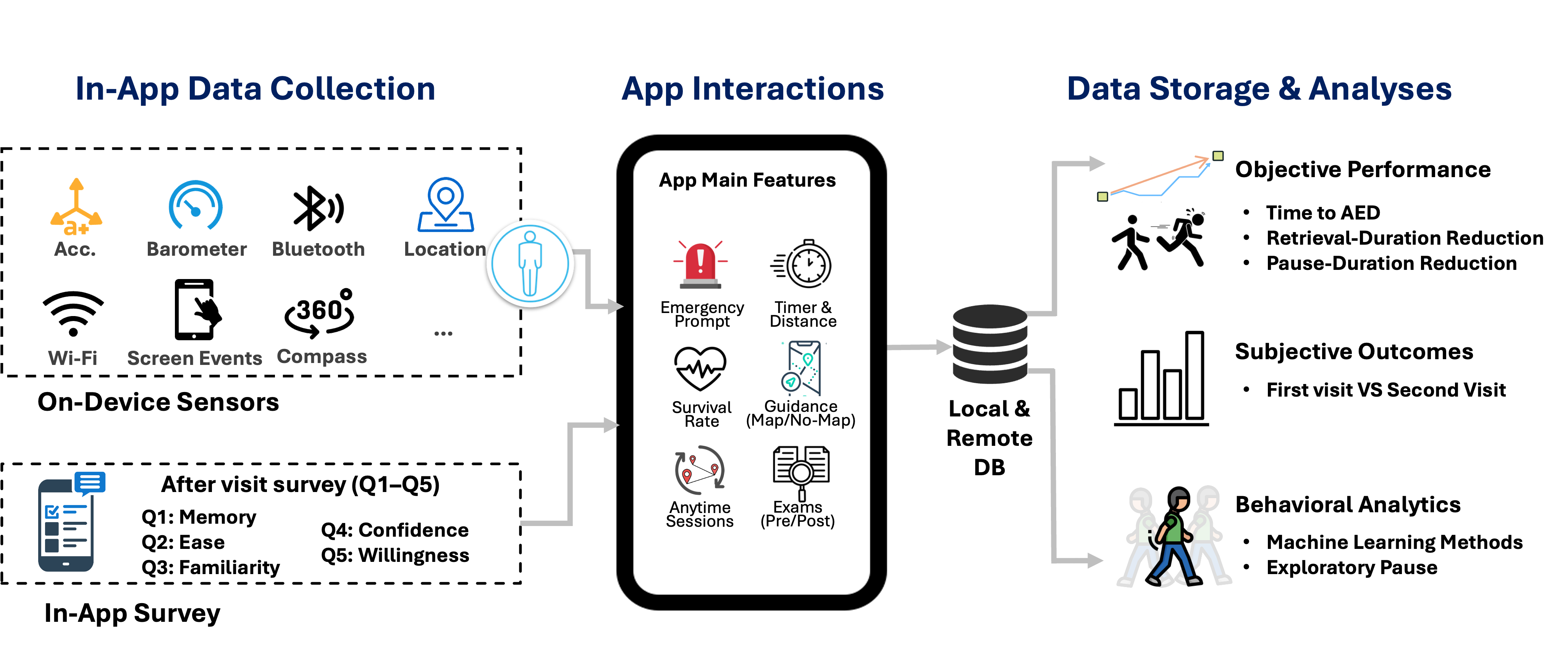}   
  \caption{AEDHunter Study Framework.}
  \Description{}
  \label{fig:framworkaed}
\end{figure}

In response to these challenges, we present AEDHunter, a gamified mobile application designed to support repeated, in-situ AED retrieval practice, with the goal of increasing bystander familiarity with AED locations and retrieval efficiency. AEDHunter integrates location-based exploration with on-device sensing and lightweight proximity verification to objectively segment retrieval attempts into distinct phases and to characterize user behavior during practice sessions. As shown in Figure~\ref{fig:framworkaed}, AEDHunter links in-app sensing and user interactions to a storage-and-analysis pipeline that produces objective performance metrics (e.g., time to AED), subjective outcomes (e.g., post-visit survey), and behavioral analytics (e.g., exploratory pauses). While prior work has often focused on the effects of gamification elements in encouraging AED visitation, AEDHunter extends beyond mere visitation by integrating in-situ practice with sensor-informed logging to analyze retrieval behaviors at the phase level.

The main contributions of our work are as follows:
\begin{itemize}
\item \textbf{Enhancing bystander knowledge of AEDs.} AEDHunter transforms AED discovery into an engaging exploration game, enabling participants to gain familiarity with nearby devices through location-based tasks and helping embed AED locations more concretely in memory.

\item \textbf{Sensor-enabled retrieval insights.} AEDHunter leverages smartphone sensors and lightweight proximity-based verification to gather detailed data on participant behaviors, clearly defining and tracking each step of the retrieval process (e.g., Preparation, Building Search, and Indoor AED Search). We implemented a two-state movement detector that distinguishes between moving and exploratory pausing, using duration of exploratory pausing as a behavioral learning signal to reveal hesitation and quantify reductions in such hesitation over time.

\item \textbf{Empirical validation.} Training reduced the group median retrieval time from 132.7 s to 97.3 s and enhanced participants’ self-reported confidence in assisting in potential OHCA events. These results suggest that gamified, sensor‑based training approaches can yield tangible benefits for time-critical AED retrieval preparedness.

\end{itemize}

\section{RELATED WORK}

Our study focuses on the use of a gamified, location-based mobile application to increase public awareness of AED locations and to understand participants' real-world retrieval behaviors through sensor data analysis. Accordingly, in this section, we review related work on gamified approaches to emergency preparedness and AED engagement, as well as advances in mobile behavior tracking technologies.

\subsection{Game-based Approaches for Emergency Preparedness and AED Engagement}
Game-based approaches have been used to enhance public engagement, motivation, and skill acquisition in emergency preparedness. Gamification, which involves adding game elements such as points, missions, and immediate feedback to non-game training tasks~\cite{deterding2011game,zainuddin2020impact}, has been reported to increase motivation, practice frequency, and retention of response protocols~\cite{kappen2013kaleidoscope,wibisono2019self,lovreglio2021comparing}. Serious games are self-contained training games with explicit rules and goals, sometimes incorporating narratives or levels~\cite{alsawaier2018effect, connolly2012systematic}. They have been used to recreate urgency and time pressure, and studies that incorporate Stress Exposure Training principles report gains in cognitive and behavioral readiness and decision-making under stress~\cite{daylamani2022framework}. Virtual and augmented reality (VR/AR) simulations serve a similar purpose, offering immediate feedback and a level of controlled situational complexity that is otherwise hard to achieve~\cite{feng2020immersive,cao2019virtual,d2023non}. Taken together, gamification, serious games, and VR/AR approaches can strengthen crisis-response capabilities and help bridge the gap between theoretical knowledge and practical application.

Furthermore, a few studies have applied similar gamified strategies specifically to AED awareness and community participation. Merchant et al.~\cite{merchant2014hidden} leveraged a web-based contest to crowdsource AED location data and encourage creative approaches for increasing AED visibility. Hussein et al. and Hao et al.~\cite{bin2019mobile, hao2019fun} developed gamified mobile applications that award points, badges, and leaderboard rankings to users who verify AED locations and their condition. While these efforts have shown promise in improving data accuracy and stimulating community engagement, they mainly focus on metrics such as user engagement or AED recall counts, rather than examining the extent to which such strategies influence real-world AED retrieval behaviors.
Complementary to these engagement-focused studies, prior work simulated AED retrieval within realistic pedestrian networks and highlighted a key design implication: public awareness of AED locations and, critically, route knowledge, particularly knowledge of the shortest path, are central to reducing intervention time~\cite{peng2024poster}. This line of work motivates the need to evaluate not only whether users ``know'' an AED exists, but whether they can translate that knowledge into faster, procedurally efficient retrieval. Building on these insights, we developed a gamified mobile application designed to incentivize users to physically visit and activate AEDs. In our preliminary report~\cite{peng2025aedhunter}, we conducted pilot experiments and collected subjective self-evaluations of preparedness. That initial evaluation primarily provided self-reported engagement data from single-campus deployments, while objective evidence of learning, procedural improvement, or formal performance metrics remained underexplored. To address this gap, we extend the application beyond simple ``visiting'' or recall measures by introducing a structured pre–post–transfer evaluation, phase-level performance attribution, and formal statistical analyses across two campuses to quantify retrieval-time improvements and identify the stages where time savings occur.

\subsection{Behavior Tracking Technologies}
Mobile phones, equipped with affordable yet powerful sensors such as accelerometers, Wi-Fi, Bluetooth, and directional sensors, enable detailed tracking of environmental dynamics and human behavior. Accelerometers, for instance, have been widely utilized to identify physical activities and daily routines~\cite{wang2014studentlife, wang2016crosscheck, zhang2022if, dong2022doublecheck, enokibori2024rtsfnet}. Wang et al.~\cite{wang2014studentlife} and Enokibori~\cite{enokibori2024rtsfnet} leveraged accelerometer data to classify activities such as walking and driving, while Ashqar et al.~\cite{ashqar2018smartphone} and Sun et al.~\cite{sun2023machine} combined accelerometer data with GPS information to detect transportation modes. Similarly, Bluetooth and Wi-Fi data have been employed to monitor human locations and social interactions by determining positions, trajectories, and proximity to others~\cite{faragher2014analysis, hong2016socialprobe, oosterlinck2017bluetooth, spachos2020ble, cipriani2021traffic, hasan2022someone}. Platforms such as MOCHA~\cite{nishiyama2022mocha} utilize Bluetooth beacons and smartphones to detect room-level locations, supporting applications like contact tracing and congestion monitoring. Additionally, Wi-Fi and Bluetooth sensing have been applied to monitor pedestrian and cyclist traffic, enabling estimation of travel routes and durations~\cite{lesani2018development}. Collectively, these methodologies facilitate comprehensive analyses of movement patterns and behavioral dynamics across diverse environments.

However, the application of these sensing technologies to time-sensitive emergency behaviors, such as AED retrieval, remains limited. Consequently, little is known about the specific movement patterns, decision points, or obstacles that individuals encounter when retrieving AEDs in real-world scenarios. 
Leveraging mobile sensing technologies in this context can yield new insights into real-world emergency behaviors, informing the design of more effective preparedness interventions.
Rather than relying on precise indoor localization, we use Wi-Fi BSSID and accelerometer data for coarse indoor-entry detection and to segment trips into Preparation, Building Search, and Indoor AED Search. A lightweight classifier detects exploratory pausing, which acts as a learning signal.

\section{MOTIVATIONS AND RESEARCH QUESTIONS}
This section presents the motivation, research questions, and our quantitative approach to measuring how repeated practice can enhance real-world AED retrievals.

\subsection{Motivations}

Despite efforts to raise public awareness of AEDs through gamification~\cite{merchant2013crowdsourcing, hao2019fun, bin2019mobile}, significant gaps remain in both knowledge and practice regarding how to understand and improve AED retrieval during emergencies.  However, existing approaches often overlook mobile sensing technologies that can monitor and analyze user behaviors, thereby missing critical insights into retrieval patterns and the environmental challenges users encounter when locating AEDs. Additionally, only a few studies have examined how behavioral changes during AED retrieval influence overall emergency preparedness.

Based on these identified challenges, we formulate four primary requirements for our study:

\begin{itemize}
    \item \textbf{Mental Memory}: Rather than relying solely on static navigation aids, which may be insufficient in time-pressure situations, we encourage participants to build mental maps by repeatedly visiting AED locations during low-stress periods. This approach aims to reduce confusion and hesitation during actual emergencies.

    \item \textbf{Spatial Awareness}: Given the limited public awareness of nearby AED locations, we incorporate gamified tasks that repeatedly guide users to these locations, reinforcing their spatial memory and familiarity with the surrounding context of these critical sites.
    
    \item \textbf{Quantitative Validation}: Few studies have quantitatively validated the real-world effectiveness of gamified AED training. Therefore, we employ repeated interventions and standardized metrics to systematically measure their impact on retrieval time and user confidence.
    
    \item \textbf{Behavioral Insights}: To pinpoint specific barriers encountered during AED retrieval, our approach integrates unobtrusive smartphone sensing (e.g., accelerometer, Wi-Fi signals) to identify momentary pauses or hesitations that may indicate user difficulties. These data can reveal particular obstacles and inform targeted design refinements that may enhance retrieval performance.
\end{itemize}

\subsection{Research Questions}

Aligned with these requirements, we propose three primary research questions (RQs):

\begin{itemize}
    \item \textbf{RQ1 (\textit{Awareness and Efficiency})}: Does repeated practice reduce the time required to locate and retrieve an AED in a real-world environment? We aim to provide quantitative evidence that repeated, gamified practice sessions lead to measurable time savings, thereby reducing confusion and hesitation commonly observed during emergencies.
    
    \item \textbf{RQ2 (\textit{Perception and Intention})}: Do repeated retrieval exercises influence users' confidence and willingness to intervene in a real emergency? Even with sufficient knowledge of AED locations, psychological barriers (e.g., fear or hesitation) may still deter bystanders. We investigate whether multiple practice sessions lead to sustained improvements in perceived confidence and willingness to assist.
    
    \item \textbf{RQ3 (\textit{Behavior})}: How can sensor data (e.g., accelerometer, Wi-Fi) reveal user behaviors (such as exploratory pauses) that affect retrieval outcomes? Micro-level hesitations can substantially impact overall retrieval time. We aim to systematically detect these exploratory pauses using sensor data, providing finer-grained insights than traditional observational or self-report methods.
\end{itemize}

\subsection{Definition of Retrieval Efficiency}

To evaluate whether and how our intervention addresses the identified requirements, we operationalize retrieval efficiency as a combination of objective performance outcomes and subjective perceptions of familiarity and confidence. Below, we first describe the quantitative measures used to assess retrieval efficiency, followed by participant-reported indicators capturing shifts in confidence, familiarity, and willingness to provide aid.

\subsubsection{Objective Metrics}
 To enable lightweight and scalable deployment, we define a learning signal that can be captured directly from on-device sensing. 
This signal, termed \emph{exploratory pauses}, refers to short interruptions in movement when users momentarily hesitate or reorient during task execution. At the behavioral level, exploratory pauses manifest as short stop-and-go patterns or brief pauses in motion trajectories. At the cognitive level, they are often associated with moments of uncertainty or information seeking and may indicate the transition from exploratory to familiarized behavior as spatial learning progresses.

Traditional fine-grained motion metrics (e.g., speed, path curvature) can also capture learning-related changes but often require indoor trajectory reconstruction, which is costly and sensitive to environmental conditions. In contrast, exploratory pauses can be detected from inertial or device-level motion data, offering a low-cost, generalizable indicator that can serve as a proxy for the evolving spatial familiarity of the user.

Building on this design, we report two complementary normalized metrics, in addition to the absolute pre–post reduction in completion time:
(i) the \emph{relative retrieval-duration reduction} $\Delta D_T$, capturing the outcome-level improvement in total task duration, and
(ii) the \emph{relative pause-duration reduction} $\Delta D_P$, quantifying the mechanism-level reduction in exploratory pauses.
For each pre/post trip, we record the total retrieval duration $D_T$ and the total pause duration $D_P$ (see Section~\ref{sec:pause-detect}), where subscripts ``pre'' and ``post'' denote the two evaluation sessions.

\paragraph{Relative retrieval-duration reduction $\boldsymbol{(\Delta D_T)}$.}

We define $\Delta D_T$ to quantify how much faster participants retrieve an AED after repeated gamified practice. Let $D_{T,\text{pre}}$ and $D_{T,\text{post}}$ denote the total retrieval durations (in seconds) before and after the intervention, respectively. Their normalized difference yields the relative retrieval-duration reduction. A larger $\Delta D_T$ indicates greater improvement in overall retrieval efficiency.
\[
 \Delta D_T = \frac{D_{T,\text{pre}} - D_{T,\text{post}}}{D_{T,\text{pre}}},
 \qquad \Delta D_T \in (-\infty, 1].
\]

\paragraph{Relative pause-duration reduction (learning signal) $\boldsymbol{(\Delta D_P)}$.}

We introduce $\Delta D_P$ to capture the reduction in exploratory pauses that accompanies learning. Let $D_{P,\text{pre}}$ and $D_{P,\text{post}}$ represent the total pause durations before and after the intervention, respectively. A positive $\Delta D_P$ indicates fewer or shorter exploratory pauses after the intervention. Compared with the relative retrieval-duration reduction, which measures the total time reduction for the entire retrieval process, the relative pause reduction provides a more fine-grained metric that focuses on behavioral markers of increased familiarity and reduced uncertainty, the key aspects of retrieval where improvement is most likely to occur.
\[
 \Delta D_P = \frac{D_{P,\text{pre}} - D_{P,\text{post}}}{D_{P,\text{pre}}},
 \qquad \Delta D_P \in (-\infty, 1].
\]

\subsubsection{Subjective Metrics}

In addition to objective performance metrics (e.g., retrieval time, pause duration), our study collects brief user self-reports after each retrieval session. These self-reports capture changes in perceived memory reliance, ease of locating the AED, device familiarity, and post-visit confidence and willingness to intervene. Detailed descriptions of these self-report measures are provided in Section~\ref{sec:key_features}.

\begin{figure*}[tb]
  \centering
  \subfigure[Initial screen]{
    \includegraphics[trim=0mm 30mm 0mm 8mm, clip, width=0.18\linewidth]{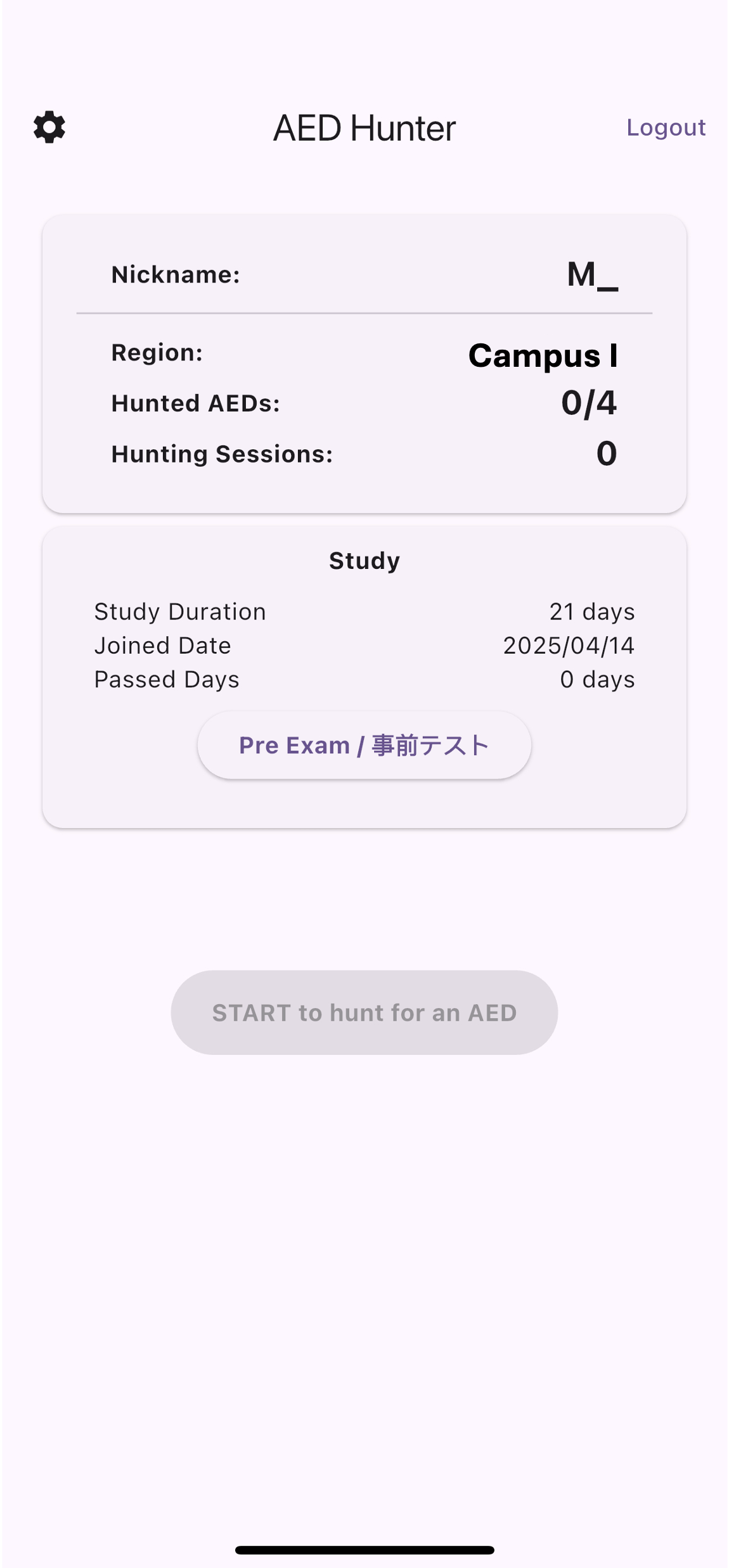}
    \label{fig:preexammain.png}}
  \subfigure[Approaching screen]{
    \includegraphics[trim=0mm 30mm 0mm 8mm, clip, width=0.18\linewidth]{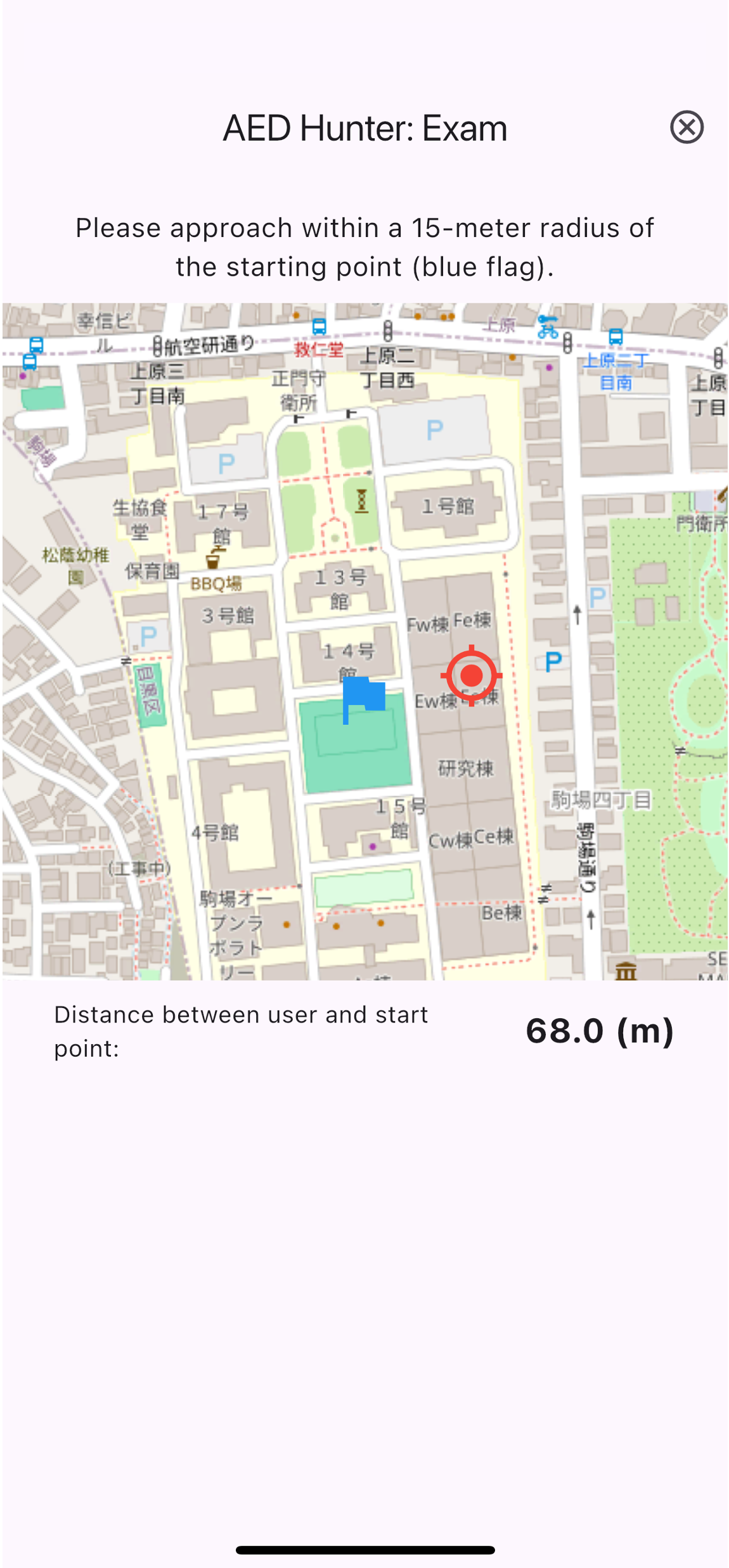}
    \label{fig:exam_approch}
  }
  \subfigure[Ready to start screen]{ 
    \includegraphics[trim=0mm 30mm 0mm 8mm, clip, width=0.18\linewidth]{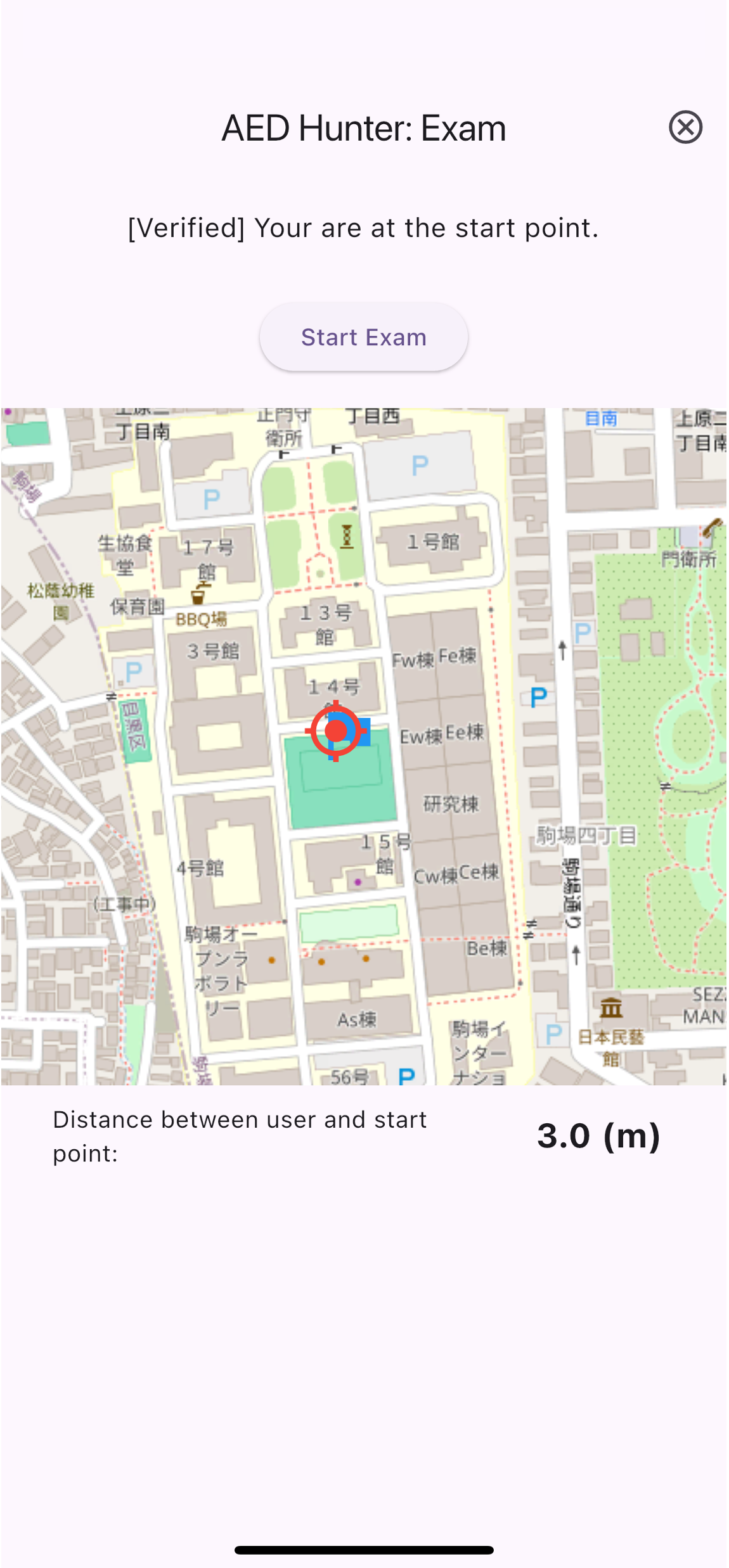}
    \label{fig:exam_start}
  }
  \subfigure[Countdown screen]{
    \includegraphics[trim=0mm 30mm 0mm 8mm, clip, width=0.18\linewidth]{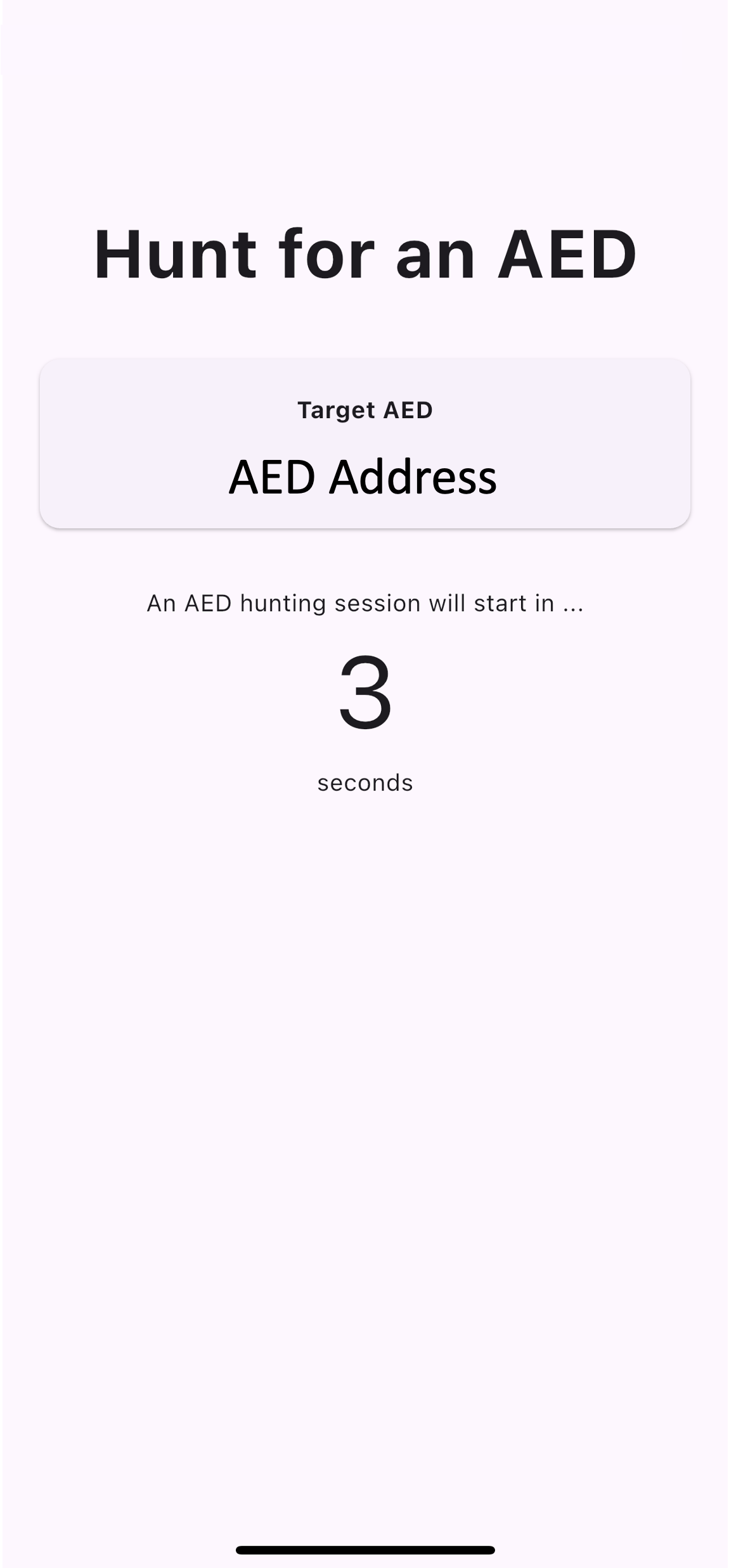}
    \label{fig:countup_screen}
  }
  \subfigure[Exam screen]{
    \includegraphics[trim=0mm 30mm 0mm 8mm, clip, width=0.18\linewidth]{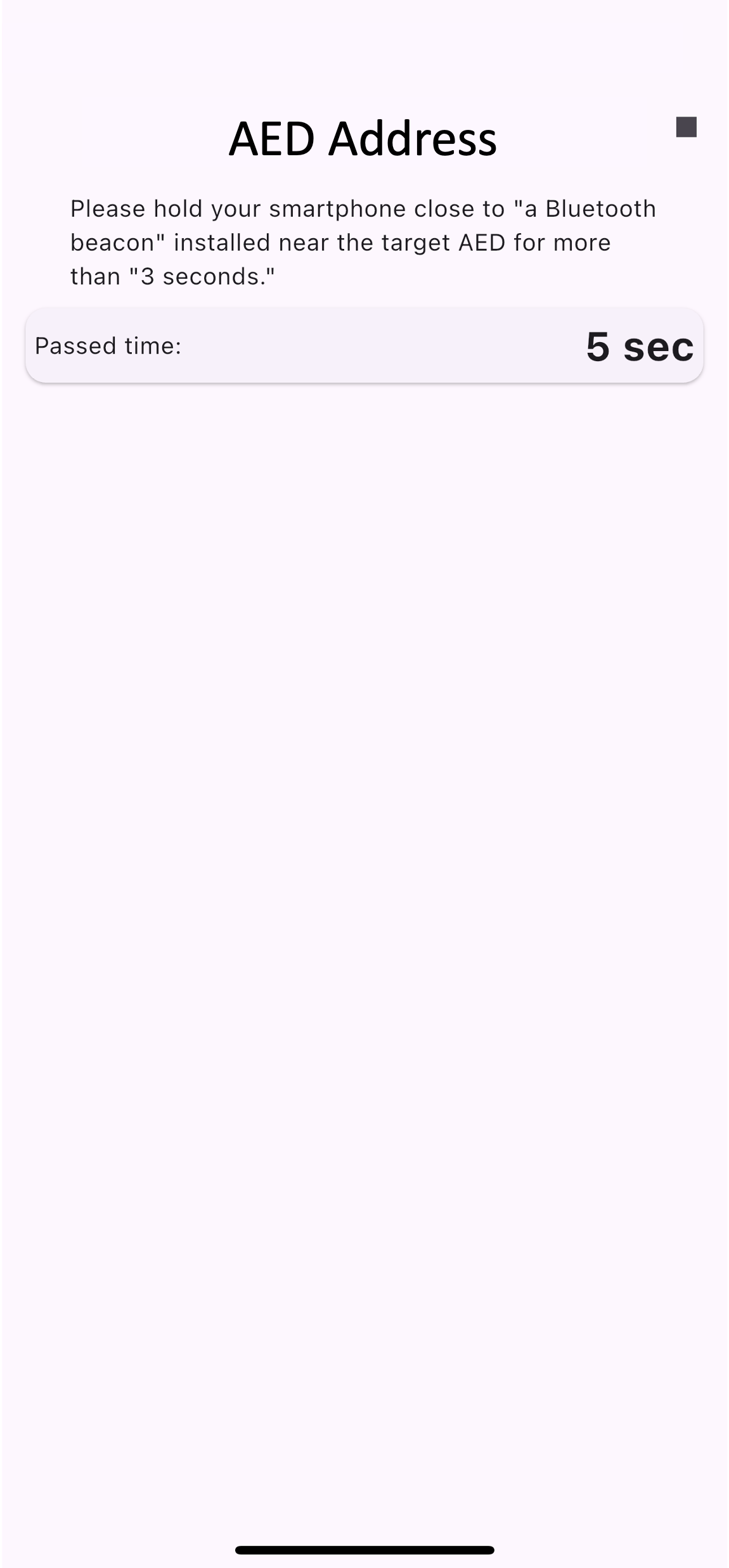}
    \label{fig:exam_on}
  }
  \caption{Screenshots of the AED hunting exam process. The process begins with the initial screen (a), where the participant's profile and study information are displayed. After selecting the hunting mode, the participant proceeds to the approaching screen (b), which shows the distance to the designated starting point on a map. Once within the required radius, the ready-to-start screen (c) confirms that the participant is correctly positioned. A countdown (d) then initiates the AED hunting session, after which the participant finds the target AED (e) by approaching the Bluetooth beacon and verifying proximity for a set duration.}
  \Description{Screenshots of the AED hunting exam process.}
  \label{fig:scr_exam}
\end{figure*}

\begin{figure*}[tb]
  \centering
  \subfigure[Main screen]{
    \includegraphics[trim=0mm 30mm 0mm 8mm, clip, width=0.18\linewidth]{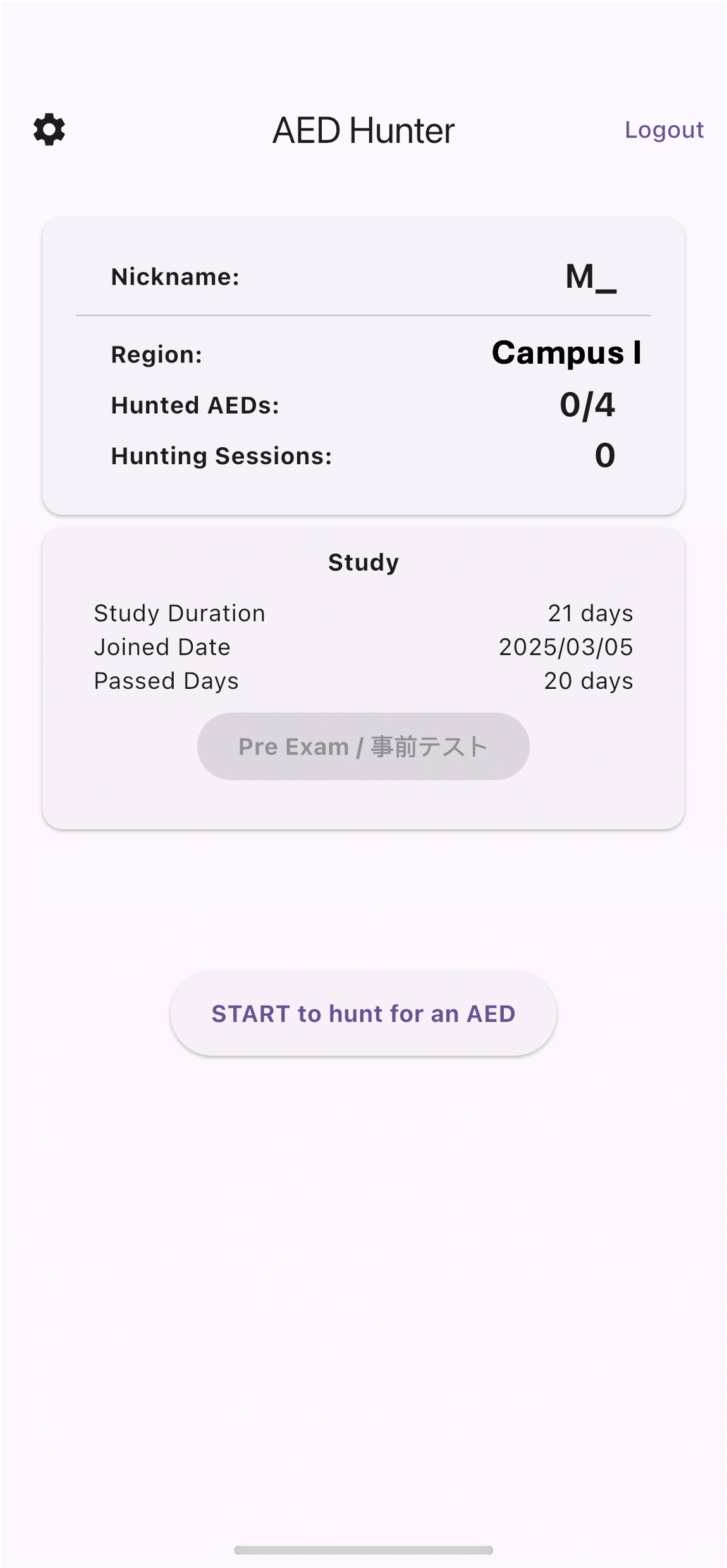}
    \label{fig:main_screen}
  }
  \subfigure[Simulation screen]{ 
    \includegraphics[trim=0mm 30mm 0mm 8mm, clip, width=0.18\linewidth]{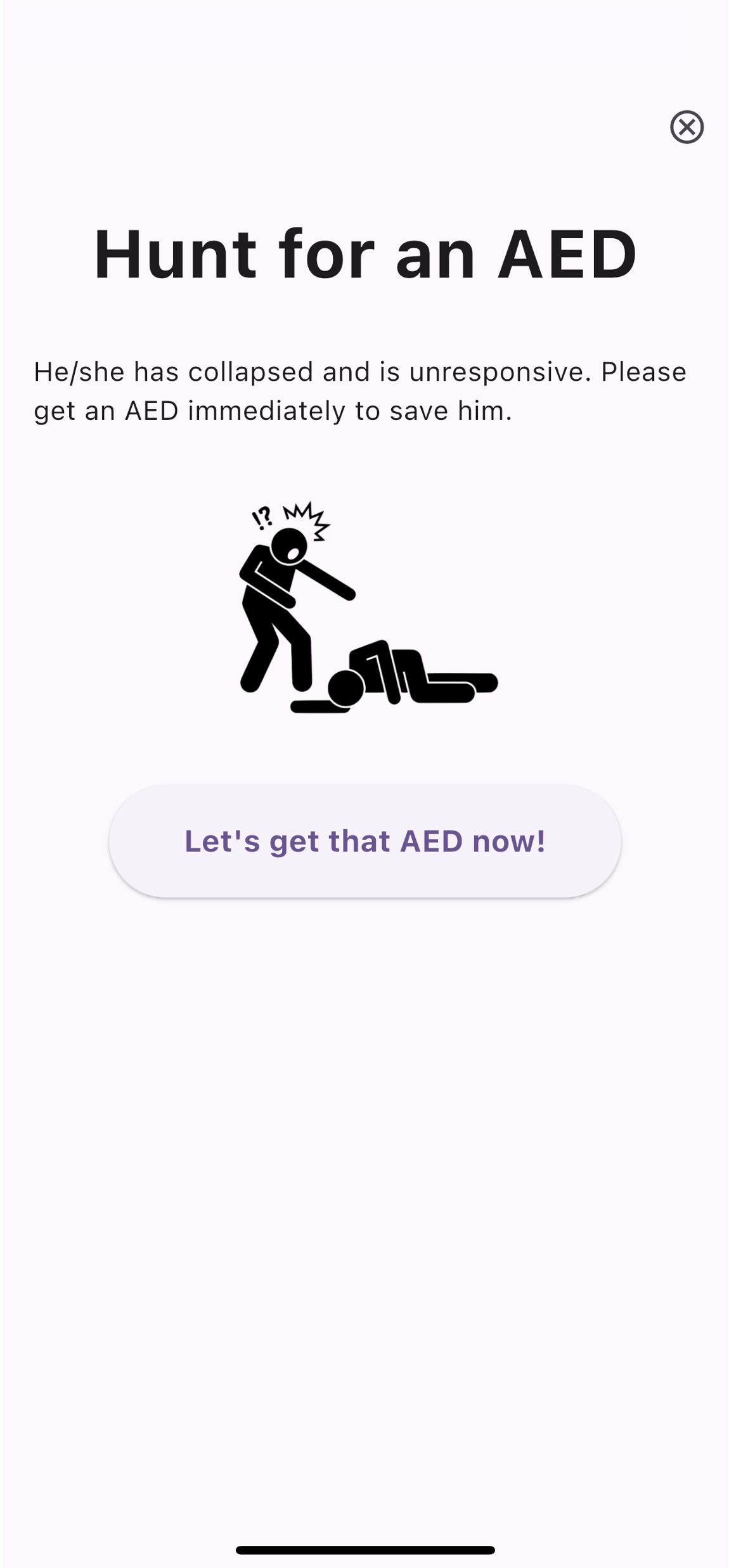}
    \label{fig:announce_screen}
  }
  \subfigure[Countdown screen]{
    \includegraphics[trim=0mm 30mm 0mm 8mm, clip, width=0.18\linewidth]{fig/infoscreen.png}
    \label{fig:infoscreen}
  }
  \subfigure[Game screen]{
    \includegraphics[trim=0mm 30mm 0mm 8mm, clip, width=0.18\linewidth]{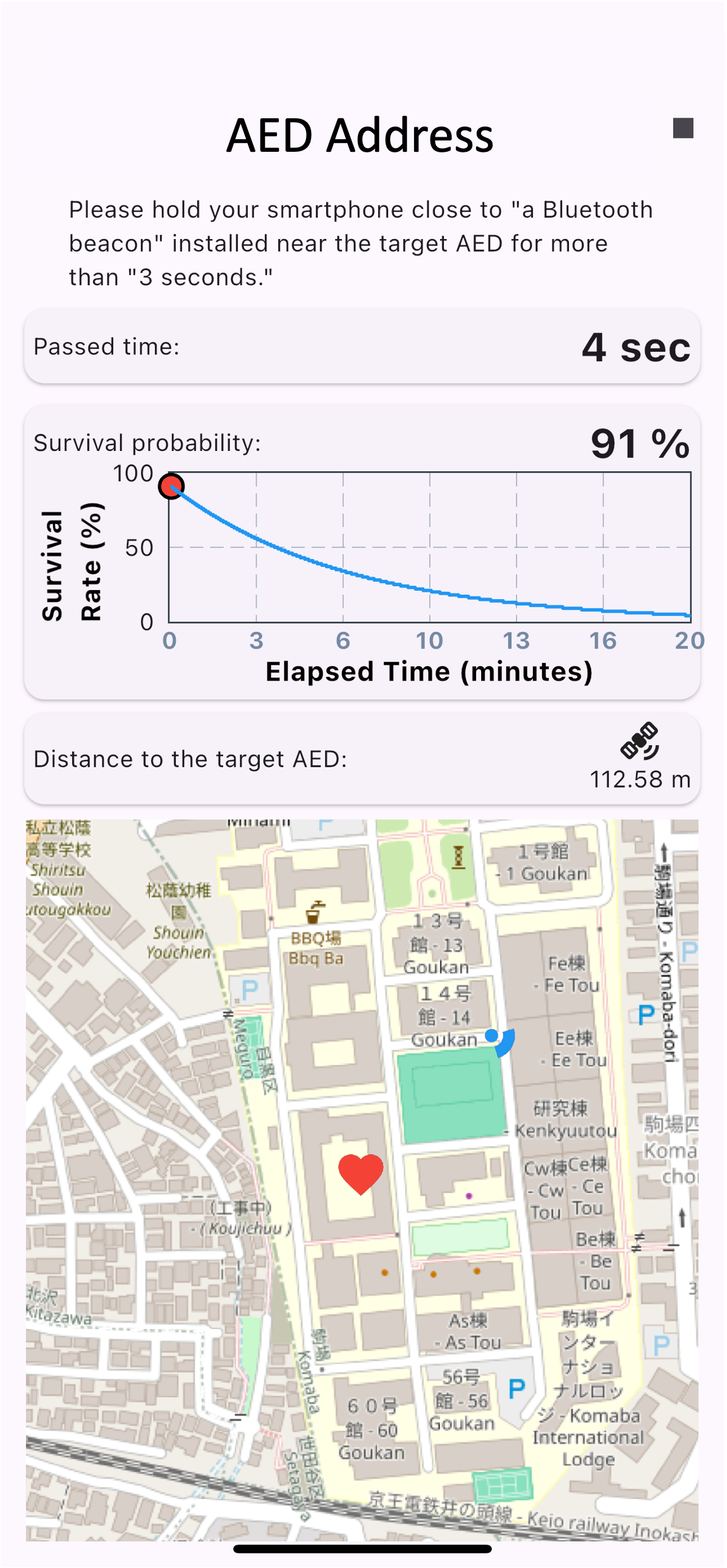}
    \label{fig:map_screen}
  }
  \subfigure[Verification screen]{
    \includegraphics[trim=0mm 40mm 0mm 0mm, clip, width=0.18\linewidth]{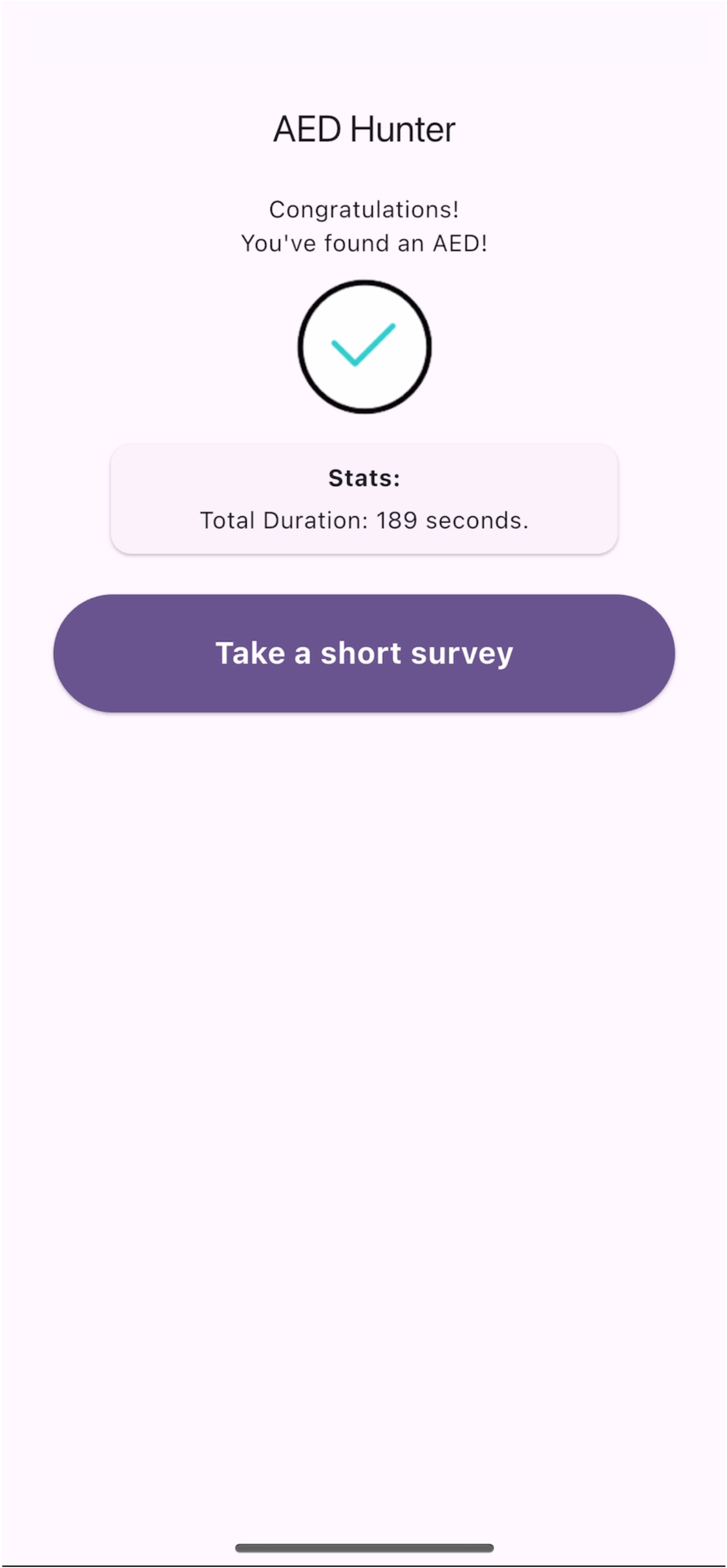}
    \label{fig:scr_genral}} 
  \caption{Screenshots of the routine game sessions.
The routine game session begins with the main screen (a), where the participant can initiate a hunting session. After starting, the simulation screen (b) displays a scenario prompting the participant to find an AED immediately. This is followed by a countdown screen (c) that indicates the imminent start of the search. During the game screen (d), the participant is guided to approach the AED location, with indicators such as signal proximity and elapsed time. Upon successfully locating the AED, the verification screen (e) confirms the completion of the session and prompts the participant to proceed to a post-session survey.}
  \Description{Screenshots of the routine game sessions.}
  \label{fig:screenshots_of_the_app}
\end{figure*}

\section{SYSTEM DESIGN: AEDHUNTER}
AEDHunter uses lightweight gamification for brief, participant-initiated in-situ practice; accordingly, our evaluation focuses on (i) retrieval efficiency (time to reach an AED), (ii) perceived confidence and willingness to assist, and (iii) sensor-derived micro-behaviors during retrieval (e.g., exploratory pauses).
Specifically, AEDHunter was designed to fulfill three primary objectives. \textbf{Goal 1—Awareness} aims to help the public discover and become familiar with nearby AED locations. \textbf{Goal 2—Retrieval Skill} seeks to build participants' ability and confidence to rapidly retrieve an AED during time-critical emergency situations. Finally, \textbf{Goal 3—Behavioral Insight} focuses on capturing fine-grained sensor and survey data to reveal how individuals move, make decisions, and learn within real-world environments.

\subsection{Design Principles}

We designed AEDHunter based on four guiding principles to create an engaging, informative, and data-rich user experience.

\textbf{(1) Flexible Access.} Participants can initiate sessions whenever and wherever they prefer, accommodating diverse schedules and encouraging spontaneous participation. Multiple starting points allow users to explore various routes to the AED, fostering familiarity with their surroundings, an essential skill for responding effectively in unexpected emergencies.

\textbf{(2) Progress Awareness.} AEDHunter displays real-time survival probability, elapsed time, and remaining distance to the target AED. These indicators emphasize the urgency of rescue efforts and highlight how each moment can be pivotal in life-saving situations, thereby simulating the time pressure and sense of urgency present in real-world emergencies.

\textbf{(3) Comprehensive Data Collection.} 
The application records detailed participant behavior through sensor logs (e.g., elapsed time, acceleration, gyroscope, and compass direction) and complements these objective measurements with concise in-app surveys capturing participant impressions. This integrated approach provides comprehensive insights into participants' movement patterns and spatial decision-making.

\textbf{(4) Iterative Practice.} 
Participants repeatedly practice retrieving the same AED under varying conditions. By coupling these tasks with immediate post-session reflections, users can reinforce their skills and build confidence, potentially improving their readiness for real-world emergencies.
\subsection{Intended Use and Deployment Scenarios}

AEDHunter is a training‑oriented, bring‑your‑own‑device mobile application that helps governments and organizations to promote public familiarity with AED locations through interactive, real‑world practice. It supports systematic deployment across various environments (e.g., campuses, office buildings, and transportation hubs) and enables brief, repeatable self-initiated practice within daily routines. This dual-layer design serves both organizational preparedness management needs and individual spatial learning objectives.

Administrators-such as education authorities, public health and emergency management agencies and enterprises-configure sites, verify and maintain AED records. They can also track coverage and performance over time. AEDHunter requires no dedicated hardware and operates using AED metadata (e.g., floor and coordinates) for configuration. Low-cost Bluetooth tags support objective arrival confirmation. During sessions, smartphones capture essential training behavioral signals through built-in sensors and post-visit surveys, enabling analysis of user behavior and familiarity.

AEDHunter supports both self‑initiated and structured use. Routine game sessions allow users to start brief, repeatable sessions from their current location. Typical examples include a student practicing on the way from the cafeteria to a classroom, an office worker starting a quick session while walking from the parking lot to the office, or a commuter checking for nearby AEDs while passing through a train station. These lightweight sessions fit naturally into daily life. 
In contrast, the built‑in exam sessions are structured activities scheduled by administrators or researchers. These exams follow standardized tasks and timing, enabling consistent assessment of learning progress and knowledge transfer across participants. The same mechanism can be adopted in real‑world deployments. For example, schools may schedule periodic AED familiarity checks as part of health‑education curricula, or workplaces may integrate similar evaluation sessions into annual safety training. In this way, AEDHunter accommodates both spontaneous practice and organized preparedness programs.

\subsection{Key Features and Operation}~\label{sec:key_features}
AEDHunter includes three primary modules: (a) a main screen where participants can register, monitor their study progress, and access study tasks; (b) routine game sessions designed to build familiarity and confidence through repeated practice; and (c) structured exam sessions (Pre-exam, Post-exam I, Post-exam II) that evaluate baseline ability, performance improvements, and the application of learned navigation skills in new locations.

\subsubsection{Main Screen} 
Figure~\ref{fig:main_screen} shows the main screen of AEDHunter, which provides participants with an overview of their study status and available actions. New participants can join the study, register, and choose their preferred location settings. Once enrolled, the screen displays cumulative progress, such as total days participated, number of hunts completed, and any pending tasks and exams. 
 Upon enrollment, the ``Pre-exam'' button is enabled so participants can assess their baseline skills; once they successfully complete the Pre-exam, that button is greyed out (disabled) and the ``Start to Hunt for an AED'' button becomes active, marking the beginning of the routine game sessions. At the end of the study period, the screen activates the ``Post-exam I'' and ``Post-exam II'' buttons, allowing participants to undertake their final evaluations.

\subsubsection{Routine Game Sessions} 
Participants can initiate these sessions anytime and anywhere at their convenience. The procedure of a routine game session is shown in Figure~\ref{fig:screenshots_of_the_app}:

\noindent\textbf{Emergency Scenario Simulation.} Participants can launch routine sessions by tapping ``Start to Hunt for an AED'', as shown in Figure~\ref{fig:main_screen}. Upon initiation, participants immediately encounter a simulated emergency scenario featuring an image of an unresponsive individual. They also see an urgent prompt: ``Someone has collapsed and is unresponsive. Please get an AED immediately to save them.'' This screen is shown in Figure~\ref{fig:announce_screen}. By tapping ``Let's get the AED now!'', participants activate the AED retrieval procedure and receive information about the target AED's location, including building name and floor, as shown in Figure~\ref{fig:infoscreen}.

\noindent\textbf{Real-Time Performance Feedback.} During the AED retrieval process, the application continuously updates the elapsed time, remaining distance, and estimated survival probability for the cardiac arrest victim, as shown in Figure~\ref{fig:map_screen}. The survival probability derives from an exponential-decay model fitted to real-world data, formally expressed as
\begin{equation}
\label{eq:survival_rate}
y = 92.13 \cdot e^{-0.147x}
\end{equation}
where $y$ denotes survival probability and $x$ is the elapsed time in minutes since cardiac arrest~\cite{holmberg2000effect}. To approximate the total time for both retrieving and delivering the AED, the application doubles the retrieval duration before calculating the survival rate. A dynamically updating graph further underscores the urgency of rescue by illustrating how survival probability diminishes with each passing second.

\noindent\textbf{Adaptive Spatial Guidance.} Context-aware cues (e.g., distance readouts and optional map hints) help participants move efficiently, aiming to reduce confusion and make them more aware of AED locations. Some participants see a map-based interface, whereas others only receive textual or distance feedback. This design flexibility allows us to explore the impact of different guidance styles on retrieval times. 
This design choice was motivated by prior research, which indicates that although external navigation aids can accelerate initial searches, they may impede the development of accurate spatial memory representations~\cite{gardony2013navigational,hejtmanek2018spatial}.

\noindent\textbf{In-App Survey and Sensor Logging.} During retrieval, smartphone sensors log movement patterns (e.g., acceleration, gyroscope, and compass direction) to derive objective behavioral measures. Immediately following each retrieval, the system launches a brief in-app survey to capture subjective responses regarding the just-completed AED retrieval, as shown in Figure~\ref{fig:scr_genral}:
\begin{enumerate}
  \item[Q1:]\label{Q1} \textit{Memory Reliance} assesses reliance on memory versus application guidance to locate the target AED in this retrieval (1: complete reliance on application guidance; 4: complete reliance on memory).
  \item[Q2:] \textit{Perceived Ease} gauges participants' perceptions of the session difficulty (1: Very Difficult; 5: Very Easy).
  \item[Q3:] \textit{Pre-Visit Familiarity} measures prior familiarity with the retrieved AED’s location, essential for evaluating its educational impact (1: Totally Unaware; 5: Super Familiar).
  \item[Q4:] \textit{Post-Visit Confidence} evaluates participants' confidence in retrieving the same AED after using the application, indicating the effectiveness of AEDHunter in enhancing emergency preparedness (1: Not Confident at All; 5: Very Confident).
  \item[Q5:] \textit{Post-Visit Assistance Willingness} assesses participants’ willingness to assist others in retrieving this AED in the future, reflecting AEDHunter's impact on fostering community readiness (1: Not Willing at All; 5: Very Willing).
\end{enumerate} 
\label{In-App Surveys}

  \subsubsection{Exams} AEDHunter includes three exams: a Pre-exam, a Post-exam I, and a Post-exam II.
Pre-exam is conducted at the beginning of the experiment. Participants start from a predefined location and locate a specific AED to establish baseline performance metrics. Post-exam I repeats the same AED retrieval task at the end of the experiment, allowing direct comparison with Pre-exam performance to evaluate participants’ improvement.
Post-exam II is also conducted after the experiment and introduces a new, unfamiliar AED to measure knowledge transfer beyond the original route. 
Figure~\ref{fig:scr_exam} illustrates the exam procedure. Participants must first approach the designated starting point (Figure~\ref{fig:exam_approch}). Once within 15 meters, the ``Start Exam'' button appears, and they can tap it to initiate the retrieval task (Figure~\ref{fig:exam_start}). The application then displays the address of the target AED, including the building name and floor (Figure~\ref{fig:countup_screen}). During retrieval, the screen displays only the elapsed time (Figure~\ref{fig:exam_on}), in contrast to the routing guidance provided during regular game sessions.
The AED used in the Pre-exam and Post-exam I is also included in general game sessions to measure the overall effectiveness and impact of repeated retrieval on participant performance. Conversely, the AED utilized in Post-exam II remains exclusive to exam conditions and is excluded from general game sessions, enabling isolated evaluation of knowledge transfer and retention.

\subsection{Application Implementation}
\label{sec:app_implementation}
AEDHunter was developed as an iOS application using Flutter (version 3.16.5), an open-source cross-platform UI development framework. To collect hardware and software sensor data on iOS, we used the AWARE Framework~\cite{nishiyama2020ios}. 
The hardware sensors include GPS for 2D locations, Wi-Fi Basic Service Set Identifier (BSSID), barometer, accelerometer, gyroscope, magnetometer, and Bluetooth. 
The 2D location, Wi-Fi BSSID, and barometer data are collected every second, and the motion sensor data (i.e., accelerometer, gyroscope, and compass direction) are collected at 100 Hz. 
We installed Bluetooth beacons near the AEDs, each of which transmits a Bluetooth Low Energy advertising packet in iBeacon format once per second. 
Screen touch events, device lock/unlock status, and application status information are collected for every event as software sensor data. 

Collected sensor data are temporarily saved in the smartphone's local storage as CSV files and uploaded to Google Cloud Storage when a network connection is available. Questionnaire responses, the target AED in the game session, and points are saved in Cloud Firestore using a dictionary data structure. 
Cloud Firestore also stores information about AED locations, including latitude, longitude, altitude, and place names. The application retrieves the latest AED information from Cloud Firestore. Map displays are implemented using OpenStreetMap via the Flutter plugin \verb|flutter_osm_plugin|\footnote{\url{https://pub.dev/packages/flutter_osm_plugin}}.

\begin{figure}[tb]
  \centering
  \includegraphics[width=0.9\linewidth]{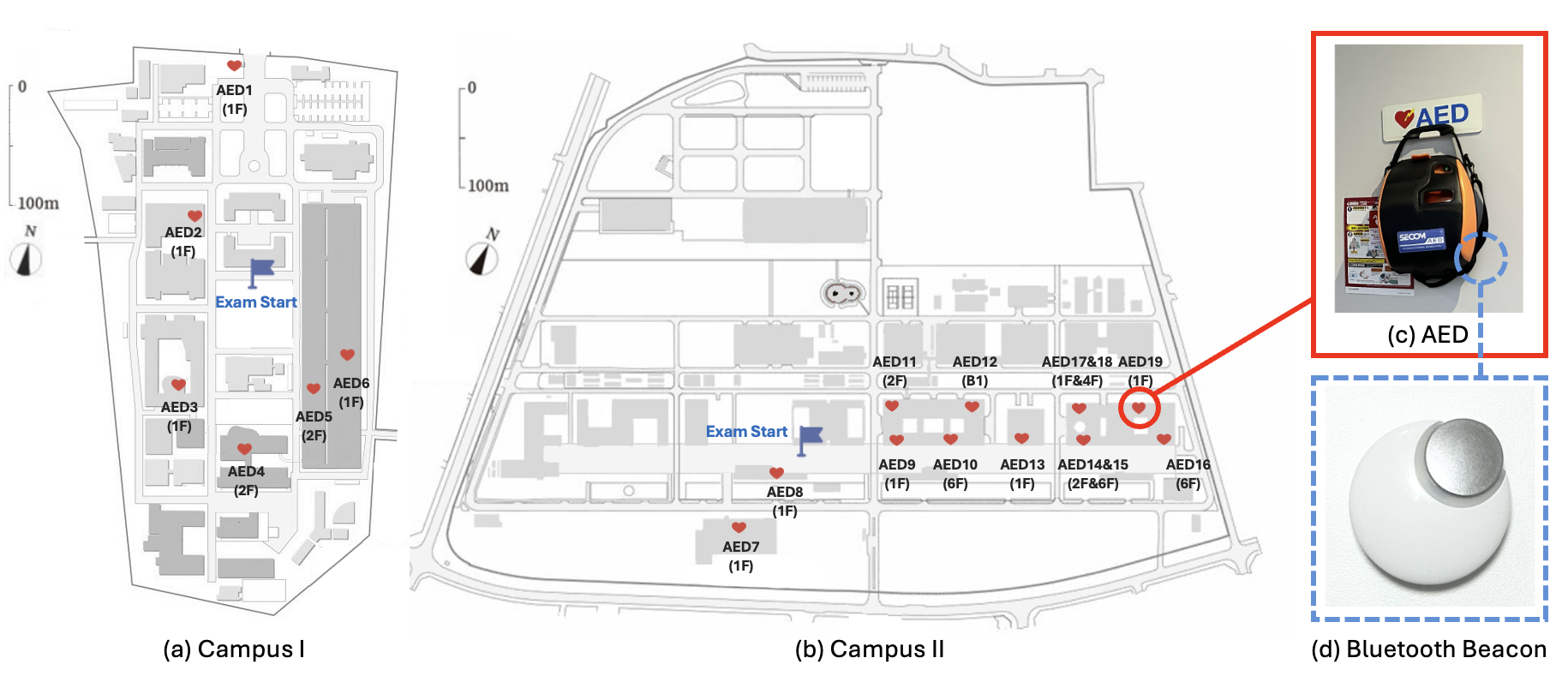} 
  \caption{Experimental Environment. (a) and (b) show maps of the two campuses, indicating AED locations (shown as hearts icons), their floor designations (e.g., ``AED 3 (2F)'' for AED 3 on the second floor), the exam‑session starting point (blue flag labeled ``Exam Start''), and pedestrian pathways (light grey lines). (c) A typical AED unit used in this study. (d) A Bluetooth beacon placed near each AED for arrival detection (each beacon measures under 3.5 cm). }
  \Description{Experimental Environment. }
  \label{fig:ex_env_detail}
\end{figure}
  
\section{EXPERIMENTS}
To evaluate the feasibility and effectiveness of AEDHunter, we conducted a user study on two university campuses after obtaining Institutional Review Board approval.

\subsection{Experimental Environment}
Figure~\ref{fig:ex_env_detail} illustrates the two university campuses where data collection took place. Campus I is equipped with six AEDs, while Campus II has thirteen. These devices are distributed across multiple buildings, from basement level 1 (B1) up to the sixth floor (6F). This distribution reflects typical AED placements in multi-building campuses, offering a realistic array of distances, floor levels, and accessibility constraints that participants might face during an emergency retrieval scenario.
We designated AED 3 on Campus I and AED 7 on Campus II for Pre-exam and Post-exam I, and we reserved AED 5 on Campus I and AED 11 on Campus II for Post-exam II.
During general game sessions, Campus I featured five AED locations (AED 1–4, 6), and Campus II featured twelve AED locations (AED 7–10, 12–19).

\subsection{Participants and Group Assignment}
We recruited a total of 20 participants from two campuses. Nine participants (5 males, 4 females; age range 22--29, $M$ = 24.5) were from Campus I, and eleven participants (8 males, 3 females; age range 23--28, $M$ = 25.1) were from Campus II. Before the experiment, all participants completed a survey (Pre-survey) that assessed (1) their familiarity with the campus layout (5-point Likert scale: 1--2 = Low, 3--5 = High), and (2) the AED locations they already knew in the experimental area.

Based on the survey results, participants were assigned to either the Map group or the No-Map group, balancing campus familiarity and prior AED knowledge. The Map group received geo-map assistance to locate the AEDs during the experiment, while the No-Map group did not receive mapping assistance. Aside from the map support variable, all other experimental settings are identical across groups. 
The detailed group assignment is as follows.
At Campus I (N=9), 4 participants reported High familiarity and 5 reported Low familiarity. We placed 2 High-familiarity and 2 Low-familiarity participants in the Map group (N=4), and 2 High-familiarity and 3 Low-familiarity participants in the No-Map group (N=5). 
At Campus II (N=11), 6 participants reported High familiarity and 5 reported Low familiarity. We placed 3 High-familiarity and 2 Low-familiarity participants in the Map group (N=5), and 3 High-familiarity and 3 Low-familiarity participants in the No-Map group (N=6). We ensured that the number of AEDs known by participants prior to the experiment was balanced across groups and evenly distributed between the two campuses.

\subsection{Procedure} \label{sec:procedure}
The study was conducted in three consecutive phases:
\begin{enumerate}
\item Pre-Survey and Pre-Exam:
At the beginning of this phase, participants completed the baseline survey and were assigned to groups as described above. They then took a Pre-exam to assess their initial ability to locate and retrieve a specific AED.
\item General Game Sessions:
Over the subsequent three-week period, participants engaged in the AEDHunter game sessions at their convenience, using their own smartphones (iOS devices). During these sessions, they were challenged to find different AEDs under simulated emergency conditions and time constraints. Depending on their group assignment, some participants had access to map support in the application during this period, while others did not. 
\item Post-Exam and Post-Survey:
At the end of the interactive phase, participants completed two sequential Post-exams. First, they revisited the same target AED as in the Pre-exam (Post-exam I), and subsequently they were required to locate an alternative AED (Post-exam II). Finally, participants completed a semi-structured survey to reflect on their experiences during the study and to provide feedback on their attitudes toward AEDs and AED retrieval after the intervention.
\end{enumerate}

\subsection{Activity Segmentation}~\label{sec:pause-detect}

To differentiate between ``exploratory pausing'' and ``moving'' states during AED retrieval, we developed a lightweight activity classifier that leverages multimodal sensor data. Considering the relatively small dataset size, we adopted three conventional machine learning methods, including Support Vector Machine (SVM)~\cite{cortes1995support}, Random Forest (RF)~\cite{breiman2001random}, and XGBoost~\cite{chen2016xgboost}, to demonstrate the effectiveness and potential of the collected sensor data.
Each model was trained and evaluated using the same train-evaluation split and a standard feature set consisting of window-level time- and frequency-domain statistics (mean, standard deviation, signal-magnitude area, spectral entropy, and dominant frequency).

\begin{figure}
    \centering
    \subfigure[Acceleration]{
        \begin{minipage}{15cm}
            \includegraphics[width=\textwidth]{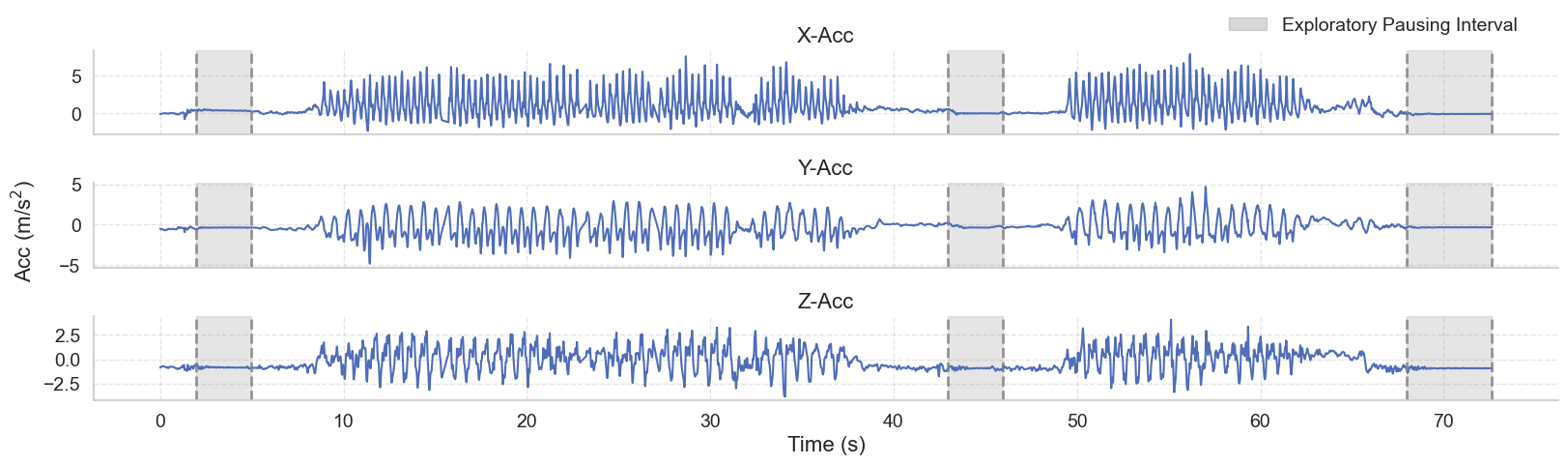}
            \label{fig:eg_acc_cal}
        \end{minipage}
    }
    \subfigure[Angular velocity]{
        \begin{minipage}{15cm}
            \includegraphics[width=\textwidth]{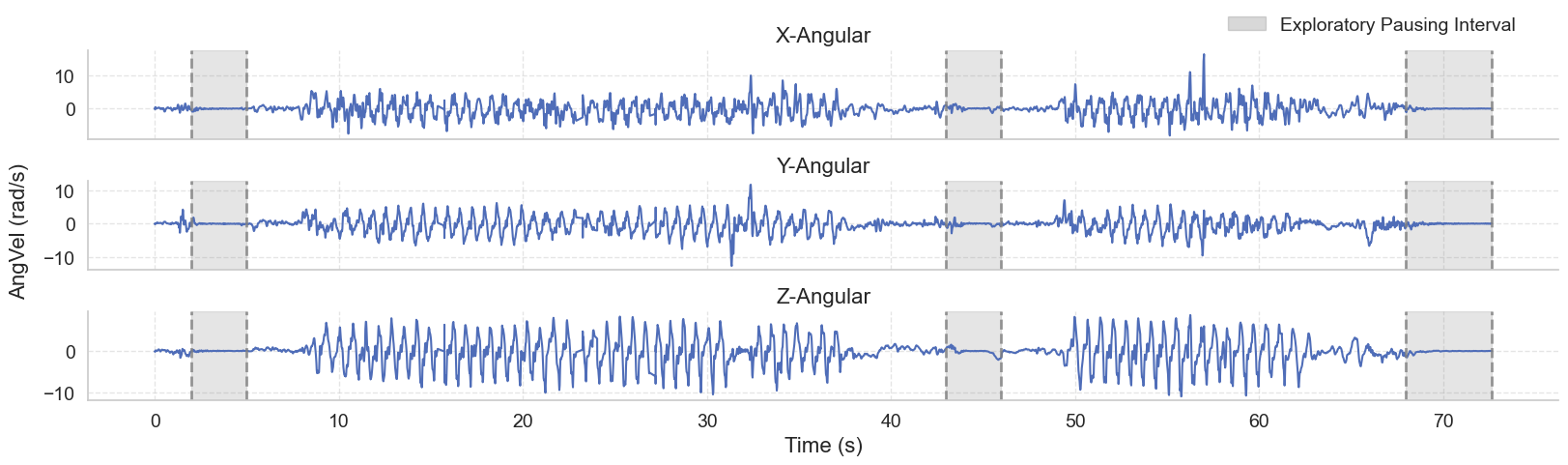}
            \label{fig:eg_grp_cal}
        \end{minipage}
    }
    \caption{Representative acceleration and angular velocity traces. Shaded (grey) regions denote segments labeled as exploratory pausing; unshaded regions correspond to segments labeled as moving.} 
    \Description{Representative acceleration and angular velocity traces. Shaded (grey) regions denote segments labeled as exploratory pausing; unshaded regions correspond to segments labeled as moving.}
    \label{fig:eg_acc_calall}
\end{figure}

We recorded 22 trips as ground truth, annotating each trip on a per-second basis as either moving (walking or running) or exploratory pausing (standing, hesitating, thinking, or checking a phone), as illustrated in Figure~\ref{fig:eg_acc_calall}. Acceleration and angular velocity data were resampled to 20 Hz and segmented into two-second intervals. In total, 1,125 segments were labeled as moving and 187 segments were labeled as exploratory pausing. To address class imbalance, we applied a data augmentation technique (SMOTE~\cite{chawla2002smote}) only to the training set. A stratified split assigned 70\% of the segments to the training set and the remaining 30\% to the evaluation set.

Among the evaluated models, the SVM demonstrated the most robust and balanced performance (weighted average F1 = 0.95), achieving superior precision and recall for both moving and exploratory pausing classes (see Appendix~\ref{appendix:b}). While Random Forest and XGBoost also delivered commendable results, their slightly lower recall and less stable performance in detecting exploratory pauses suggested a marginal disadvantage compared to the SVM. Therefore, we selected the SVM as the most effective classifier for this application, as it provided the best combination of sensitivity and specificity across both behavioral states.

\begin{figure*}[tb]
  \centering
  \subfigure[GPS trajectory]{
    \includegraphics[width=0.24\linewidth]{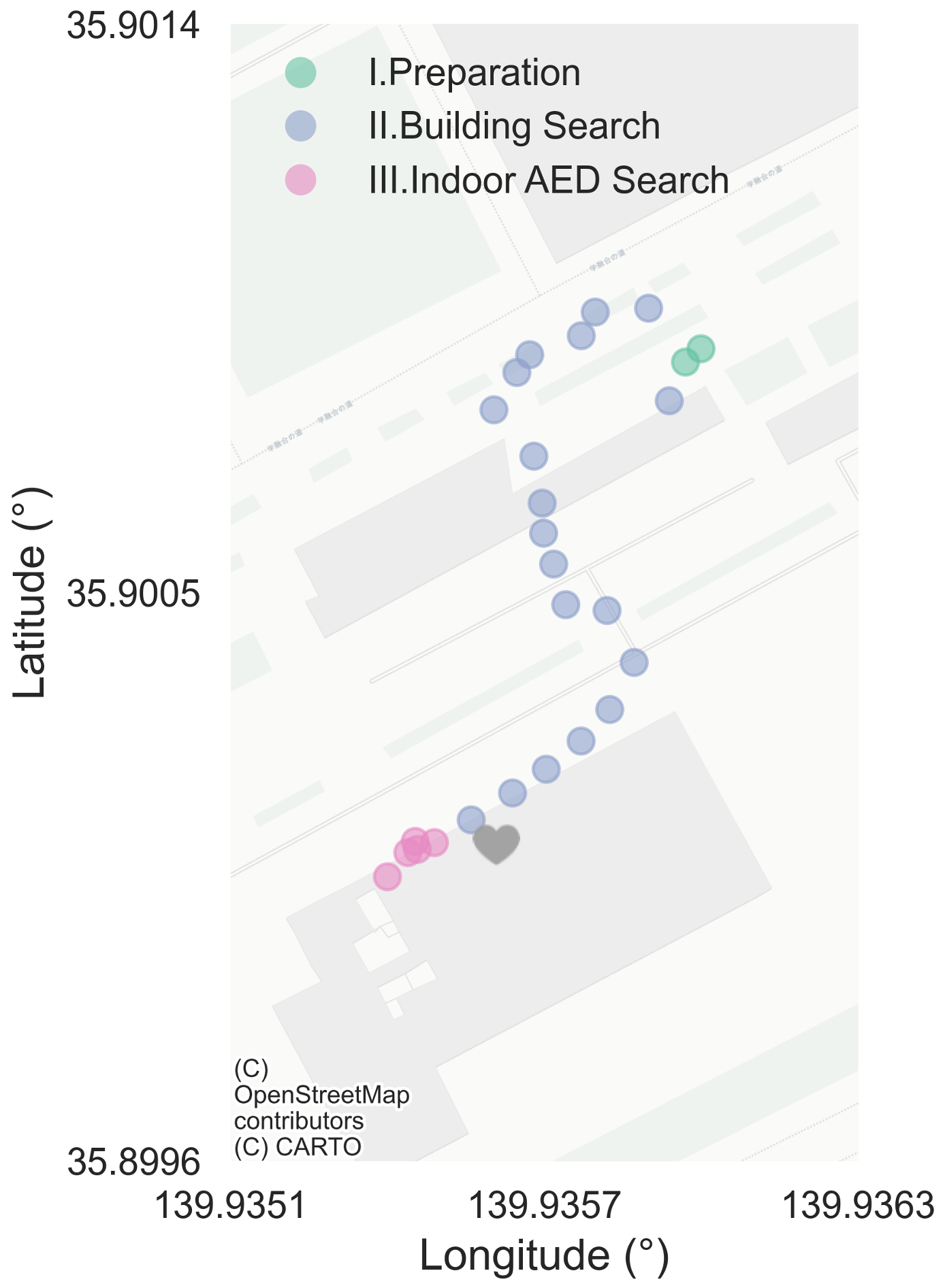}
  }\hspace{-0.02\linewidth}
  \subfigure[Sensor data]{
    \includegraphics[width=0.74\linewidth]{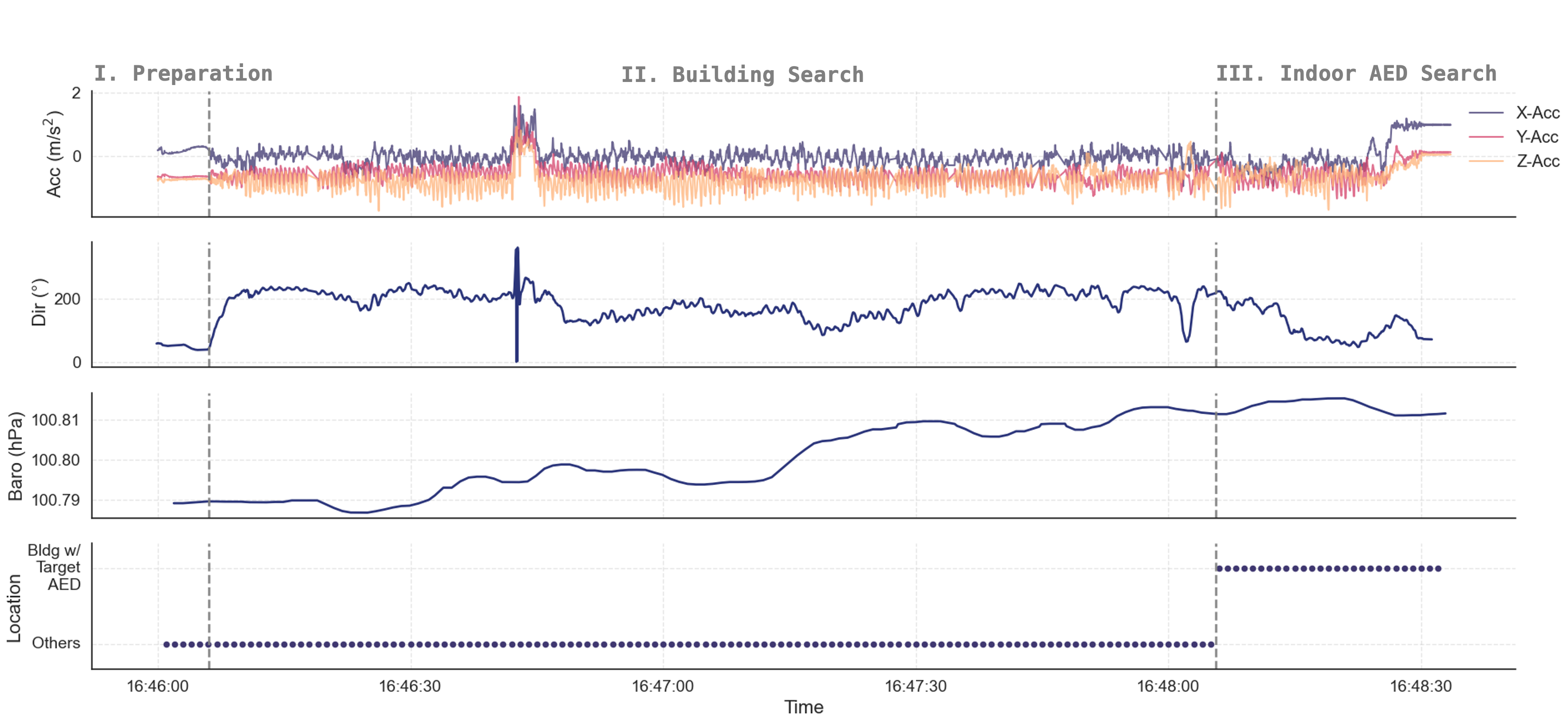}
    \label{fig:ig_acc1}}
    \caption{Example of collected data, including (a) a GPS trajectory that maps the participant's movement path and (b) corresponding sensor data such as acceleration (Acc), direction (Dir), barometric pressure (Baro), and building information derived from Wi-Fi BSSID. These data illustrate the different phases of AED retrieval activity and the duration of each phase. }
    \Description{Example of collected data, including (a) a GPS trajectory that maps the participant's movement path and (b) corresponding sensor data such as acceleration (Acc), direction (Dir), barometric pressure (Baro), and building information derived from Wi-Fi BSSID. These data illustrate the different phases of AED retrieval activity and the duration of each phase. }
  \label{fig:ig_acc}
\end{figure*}

\section{RESULTS}
In this section, we present the results of our study in five parts: (1) an overview of AED retrieval trips, (2) exam performance, (3) insights from routine game sessions, (4) survey outcomes before and after the experiment, and (5) between-group improvement analysis.

\subsection{Overview of retrieval trips}
Participants collectively conducted 228 AED retrieval trips. The median retrieval time for these trips was 124.5 seconds (IQR = 78.0–195.8 s).
Figure~\ref{fig:ig_acc} shows representative sensor data (GPS, accelerometer, Wi-Fi, and iBeacon) used to segment each trip into three stages:
\begin{enumerate} 
\item \textbf{Preparation:} An initial period, when present, characterized by low acceleration values, indicating that the participant remained largely stationary according to sensor readings, possibly due to activities such as reviewing maps or planning the route. Not all trips included this stage.
  \item \textbf{Building Search:} From the end of Preparation until entry into the target building. In this phase, participants walk toward and identify buildings equipped with the target AED.
  \item \textbf{Indoor AED Search:} From building entry until the AED is located, as detected by Wi-Fi BSSID indicating indoor presence.
\end{enumerate}

\subsection{Exam Performance}

\begin{figure}[tb]
  \centering
  \includegraphics[width=0.95\linewidth]{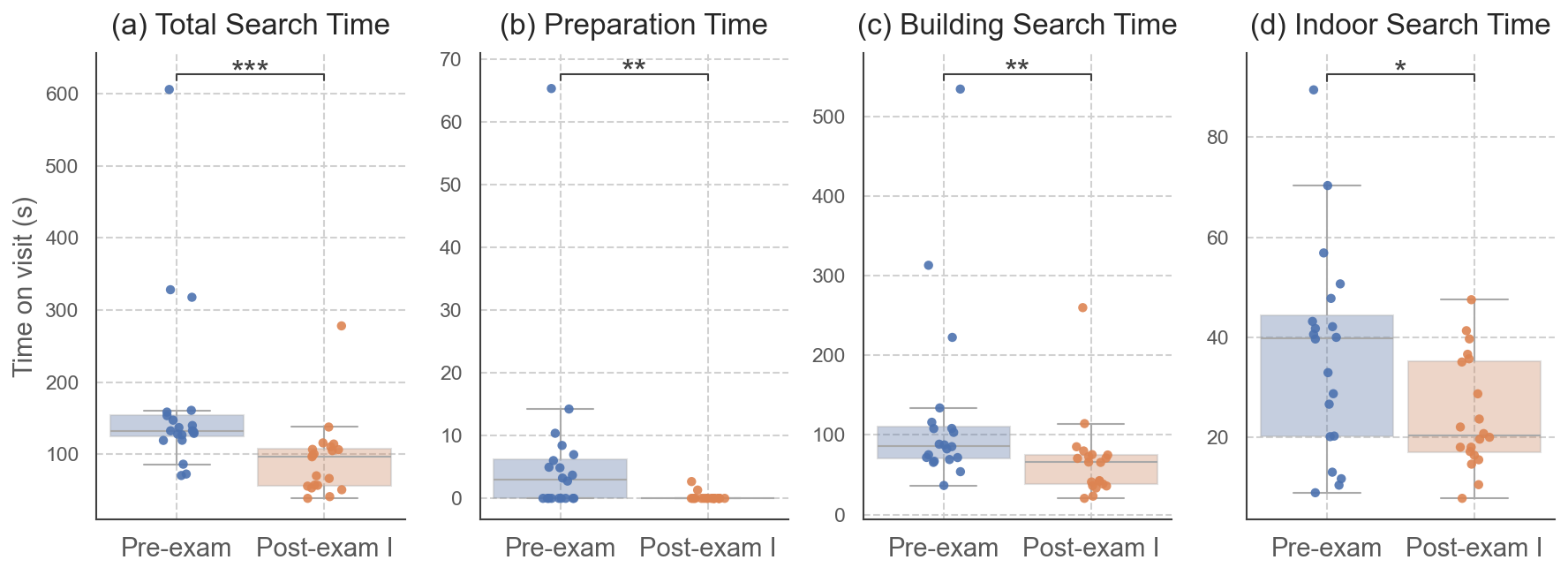}  
  \caption{
Boxplots of retrieval duration (seconds) for all participants during Pre‑exam and Post‑exam I, with individual data points shown. The figure presents (a) total retrieval time per trip and the time spent in (b) Preparation, (c) Building Search, and (d) Indoor AED Search. Brackets mark within-participant Wilcoxon signed-rank tests (one-tailed), testing for a decrease from Pre-exam to Post-exam I. (a) shows the unadjusted primary $p$, while stage‑level measures (b–d) are Holm‑adjusted within the stage family. Asterisks
indicate statistical significance: *$p < 0.05$, **$p < 0.01$, ***$p < 0.001$.}
  \Description{Box plots of retrieval duration during Pre-exam and Post-exam I }
  \label{fig:prepost_sig}
\end{figure}

We evaluated participants’ AED retrieval performance across three exams: 
before the game-based intervention (Pre-exam), after intervention (Post-exam I), and a transfer test involving unfamiliar AEDs (Post-exam II). 
Analyses were conducted in four layers to capture both outcomes and underlying mechanisms: 
(i) reduction in retrieval duration from Pre-exam to Post-exam~I, 
(ii) relative retrieval-duration reduction ($\Delta D_T$), 
(iii) behavioral learning signal quantified by the relative pause-duration reduction ($\Delta D_P$), and 
(iv) transfer performance on unfamiliar AEDs.
This structure allows us to examine not only whether participants became faster, but also how their spatial behaviors and learning processes developed through repeated practice.

\subsubsection{Reduction in Retrieval Time.} 
We first examined whether retrieval time decreased from Pre-exam to Post-exam~I at two levels: 
(i) total retrieval time, analyzed for all participants as well as separately for the Map and No-Map groups; (ii) phase-specific retrieval time, analyzed for all participants and for each group across the three task phases (Preparation, Building Search, Indoor AED Search). 
One-tailed paired Wilcoxon signed-rank tests were applied for each comparison~\cite{wilcoxon1945individual}. The null hypothesis ($H_0$) was that the within-participant median change in retrieval time equals zero, and the alternative ($H_a$) was that the median change is less than zero (i.e., retrieval time is shorter at the Post-exam~I visit). The overall test was designated as the primary analysis and was not adjusted; secondary comparisons were corrected using the Holm method to control the family-wise error rate~\cite{holm1979simple}. Figure~\ref{fig:prepost_sig} visualizes Pre-exam and Post-exam~I times for all participants, and Table~\ref{tab:sig_final} summarizes the corresponding statistics by condition and phase. 
We report the Hodges–Lehmann (HL) estimator, which represents the median of the paired differences between Post-exam~I and Pre-exam, where negative values indicate reductions. Two-sided 95\% confidence intervals (CIs) are also provided.

\textbf{Total retrieval time.} 
Pooling all participants ($N=20$), the median total retrieval time decreased from 132.7 s to 97.3 s at the Post-exam~I visit, corresponding to a HL median reduction of $-45.6$ s (95\%~CI~$[-110.3, -26.5]$, $p<0.001$). 
Using Holm-adjusted, one-tailed paired Wilcoxon signed-rank tests, both groups showed significant reductions in retrieval time
(Map group: HL~$-41.2$\,s, $p_{\text{adj}}<0.05$; 
No-Map group: HL~$-48.1$\,s, $p_{\text{adj}}<0.01$), 
indicating a statistically significant improvement in retrieval efficiency.

\textbf{Phase-specific retrieval time.} 
Across all participants, retrieval times decreased consistently across phases: 
Building Search (HL~$-27.1$\,s, 95\%~CI~$[-72.4, -16.2]$), 
Indoor AED Search (HL~$-8.6$\,s, 95\%~CI~$[-21.3, 0.0]$), 
and Preparation (HL~$-3.0$\,s, 95\%~CI~$[-6.2, -1.2]$). 
When examined separately by condition, Building Search time shortened significantly only in the No‑Map group, whereas Preparation time decreased significantly in the Map group. 
Overall, repeated task exposure primarily enhanced spatial search efficiency.

\begin{figure}
    \centering
    \subfigure[Pre-exam]{
        \begin{minipage}{12cm} 
            \includegraphics[width=\textwidth]{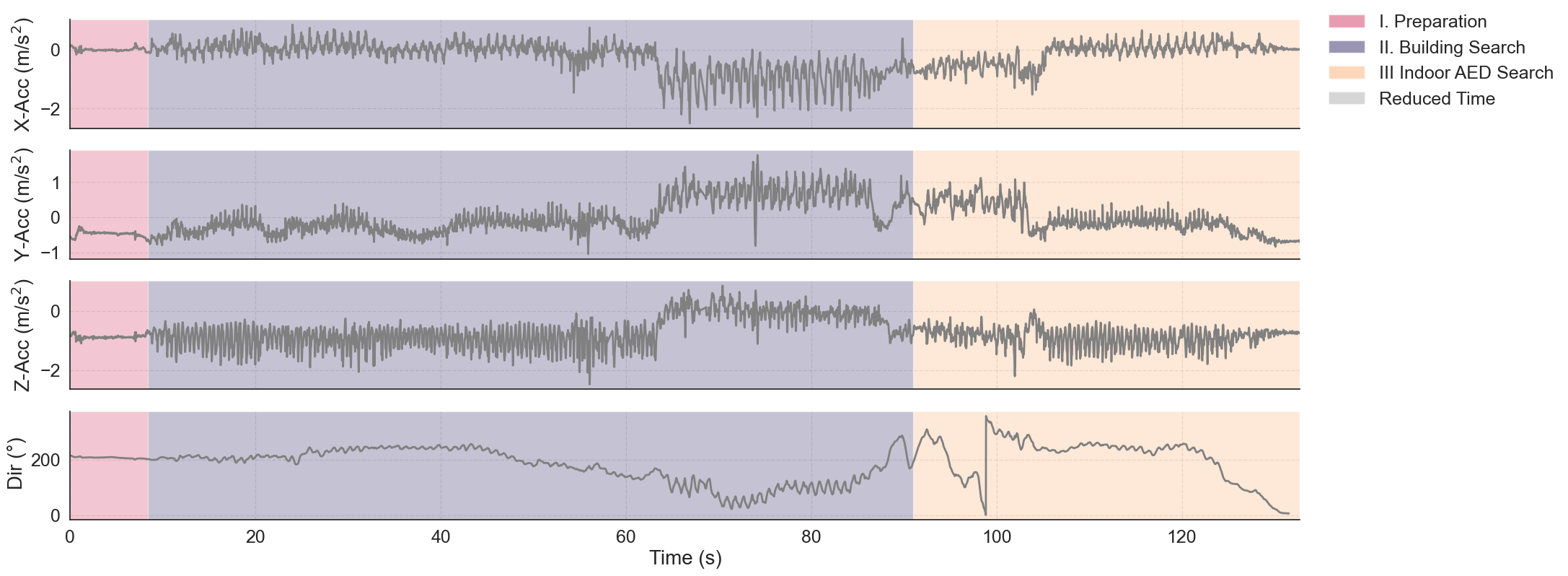} \\
            \label{fig:pre_post1}
        \end{minipage}
    }
    \subfigure[Post-exam]{
        \begin{minipage}{12cm}
            \includegraphics[width=\textwidth]{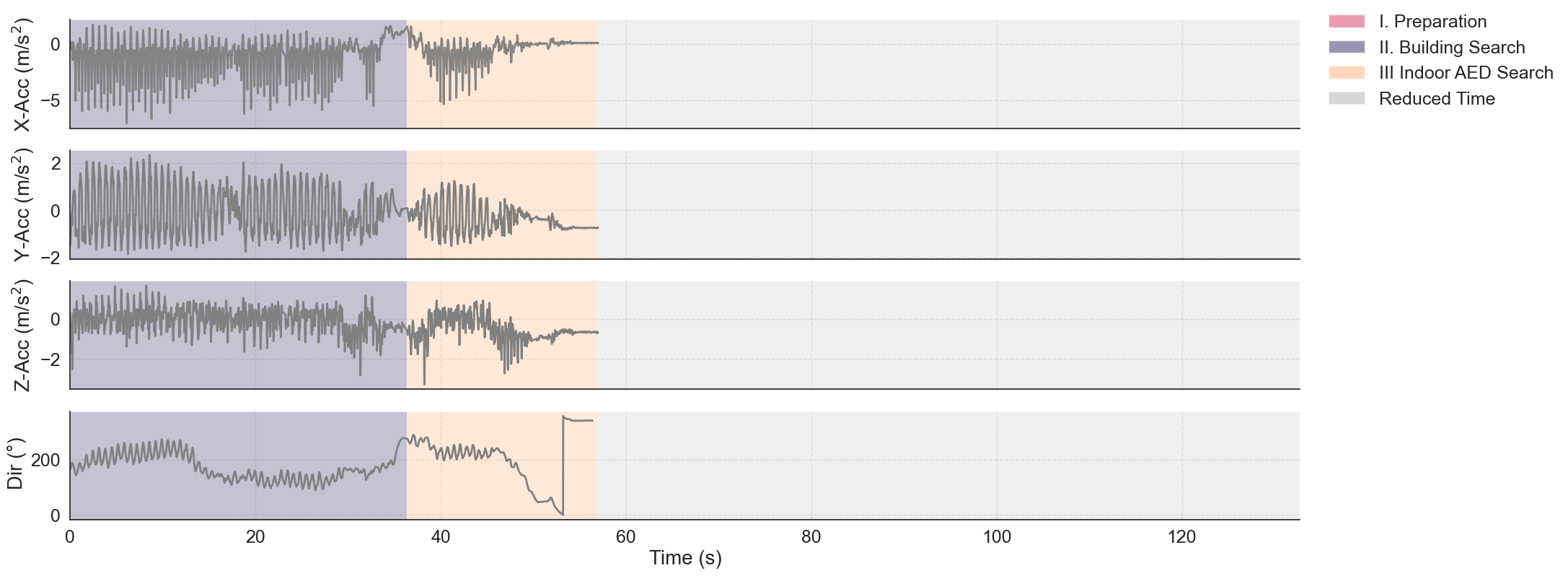} \\
      \label{fig:pre_post2}
            
        \end{minipage}
    }
  \caption{Comparison of visit times between the Pre-exam and Post-exam. Different colors indicate individual stages of the AED retrieval process, while lighter grey in the Post-exam represents the reduction in visit time for each stage.} 
  \Description{Comparison of visit times between the Pre-exam and Post-exam. Different colors indicate individual stages of the AED retrieval process, while lighter grey in the Post-exam represents the reduction in visit time for each stage.}
  \label{fig:pre_post}
\end{figure}

To visually illustrate these quantitative findings, Figure~\ref{fig:pre_post} presents exemplar retrieval trajectories. In the Pre-exam example (Figure~\ref{fig:pre_post1}), all search phases are prolonged. In contrast, the Post-exam trajectory (Figure~\ref{fig:pre_post2}) shows sharply reduced Building Search and Indoor AED Search times, with the Preparation phase eliminated, further supporting a substantial learning effect.

\begin{table}[tb]
\centering
\caption{Results of one-tailed Wilcoxon signed-rank tests. We test whether Post-exam~I completion times are lower than Pre-exam times (null: no median change; alternative: median(Post–Pre) $< 0$). Significance uses stars for significant effects (* $< 0.05$, ** $< 0.01$, *** $< 0.001$). \textit{Note:} The first row (All–Total) reports the unadjusted $p$ value; all other $p$ values are adjusted. Units: seconds.}
\label{tab:sig_final}
\fontsize{8.8pt}{10pt}\selectfont
\begin{tabularx}{\linewidth}{llrrrll}
\toprule
\textbf{Group} & 
\textbf{Scope / Phase} &
\shortstack{\textbf{Pre-exam}\\\textbf{Median Time (IQR)}} &
\shortstack{\textbf{Post-Exam I}\\\textbf{Median Time (IQR)}} &
\shortstack{\textbf{Median}\\\textbf{Reduction (HL)}} &
\shortstack{\textbf{95\% CI of}\\\textbf{Reduction}} &
\shortstack{\textbf{Adj.} \\$\boldsymbol{p}$}\\
\midrule
\multicolumn{7}{l}{\textbf{Totals}}\\
All &Total& 132.7 (125.0, 154.8) & 97.3 (56.7, 107.7) & -45.6 & [-110.3, -26.5] & *** \\
Map &Total& 147.3 (129.0, 158.5) & 98.1 (57.3, 114.2) & -41.2 & [-138.6, 6.0] & * \\
No-Map &Total& 130.7 (123.1, 138.5) & 96.5 (56.4, 106.6) & -48.1 & [-151.9, -21.9] & **\\
\midrule
\multicolumn{7}{l}{\textbf{Phase-level}}\\
All & Preparation & 3.0 (0.0, 6.3) & 0.0 (0.0, 0.0) & -3.0 & [-6.2, -1.2] & ** \\
All & Building Search & 86.5 (71.0, 110.2) & 65.8 (38.0, 75.0) & -27.1 & [-72.4, -16.2] & ** \\
All & Indoor AED Search & 39.8 (20.2, 44.3) & 20.4 (17.0, 35.2) & -8.6 & [-21.3, 0.0] & * \\
Map & Preparation & 4.9 (2.8, 6.0) & 0.0 (0.0, 0.0) & -4.9 & [-32.7, -2.4] & * \\
Map & Building Search & 87.5 (69.2, 103.3) & 65.6 (41.0, 75.4) & -23.8 & [-94.5, 44.1] & 0.098\\
Map & Indoor AED Search & 40.0 (26.6, 42.1) & 18.0 (15.5, 36.6) & -8.1 & [-36.9, 3.5] & 0.125\\
No-Map & Preparation & 0.0 (0.0, 5.1) & 0.0 (0.0, 0.0) & -1.6 & [-5.2, 0.0] & 0.128 \\
No-Map & Building Search & 82.6 (71.7, 112.1) & 66.0 (37.4, 73.9) & -31.9 & [-142.3, -12.7] & ** \\
No-Map & Indoor AED Search & 32.9 (20.2, 45.5) & 20.8 (18.4, 31.9) & -9.0 & [-22.0, 5.6] & 0.128\\  
\bottomrule
\end{tabularx}
\end{table}

\subsubsection{Relative Retrieval-Duration Reduction $\boldsymbol{(\Delta D_T)}$.}

As summarized in Table~\ref{tab:deltaT}, we examined participants’ relative retrieval-duration reduction ($\Delta D_T$) between the Pre-exam and Post-exam~I sessions. A positive $\Delta D_T$ indicates faster retrieval after the intervention, reflecting improved efficiency through reduced search time or more direct navigation. 
Across all participants ($N=20$), the median $\Delta D_T$ was 0.39, and 85\% of participants showed improvement, suggesting that repeated retrieval practice led to measurable performance gains. Both the Map ($N=9$) and No-Map ($N=11$) conditions exhibited positive median and IQR values, indicating that the efficiency improvement was not limited to a particular interface or support condition.

\begin{table}[htb]
\centering
\caption{Descriptive statistics of relative retrieval-duration reduction ($\Delta D_T$) by condition. 
Counts and percentages refer to \textbf{participants}. 
A positive $\Delta D_T$ indicates faster post-intervention retrieval. 
$N$ denotes the total number of participants per group, and $n$ denotes the number of participants in each category.}
 \label{tab:deltaT}
\begin{tabular}{lrrrrrr}
\hline
\toprule
\bf{Group} & \bf{N}  & \textbf{Median }$\boldsymbol{\Delta D_T}$\textbf{ (IQR)} & \textbf{Improved n/N (\%)} & \textbf{Worsened n/N (\%)} \\
\midrule
All    & 20  & 0.39 (0.18-0.57) & 17/20 (85.0\%) & 3/20 (15.0\%) \\
Map    & 9 & 0.36 (0.22-0.43) & 8/9 (88.9\%)  & 1/9 (11.1\%) \\
No-Map & 11  & 0.45 (0.18-0.57)  & 9/11 (81.8\%)   & 2/11 (18.2\%)  \\
\hline
\end{tabular}
\end{table}

\subsubsection{Relative Pause-Duration Reduction (Learning Signal) $\boldsymbol{(\Delta D_P)}$.}

The median (IQR) relative pause-duration reduction from Pre-exam to Post-exam~I was $\Delta D_P = 0.40$ ($-0.10$, $0.78$), suggesting a typical 40\% decrease in exploratory pauses, though with substantial inter-participant variability, including some negative values.
For inference, hesitation was operationalized as the total pause duration $D_P$ (see Section~\ref{sec:pause-detect}). We tested whether hesitation decreased from Pre-exam to Post-exam~I using a one-tailed paired Wilcoxon signed-rank test on the within-participant difference $d = D_{P,\text{post}} - D_{P,\text{pre}}$. The hypotheses were $H_{0}\!:\operatorname{median}(d) \ge 0$ (no decrease) versus $H_{1}\!:\operatorname{median}(d) < 0$ (decrease). The test indicated a statistically significant reduction in hesitation ($p < 0.05$).
Note that $\Delta D_P$ summarizes proportional change, whereas the Wilcoxon signed-rank test is based on absolute pause durations; the two are directionally consistent because $\mathrm{sign}(\Delta D_P) = \mathrm{sign}(D_{P,\text{pre}} - D_{P,\text{post}})$. Taken together, these results indicate that participants became more efficient during retrieval, with shorter pauses reflecting increased familiarity and reduced uncertainty about AED locations.

\subsubsection{Transfer Performance (Post-Exam II)}
 To evaluate knowledge transfer, participants were tasked with retrieving previously unseen AEDs during the Post-exam II session. 
 In Campus I, the median (IQR) retrieval time was 129.0 s (52.0) for the No-Map group and 106.5 s (8.3) for the Map group. 
 In Campus II, the median (IQR) retrieval time was 150.0 s (72.0) for the No-Map group and 162.0 s (81.0) for the Map group. 
 Across both campuses, median retrieval times and IQRs were comparable between groups.

\subsection{Insights from Routine Game Sessions}

\begin{figure}[tb]
  \centering
  \includegraphics[width=1\textwidth]{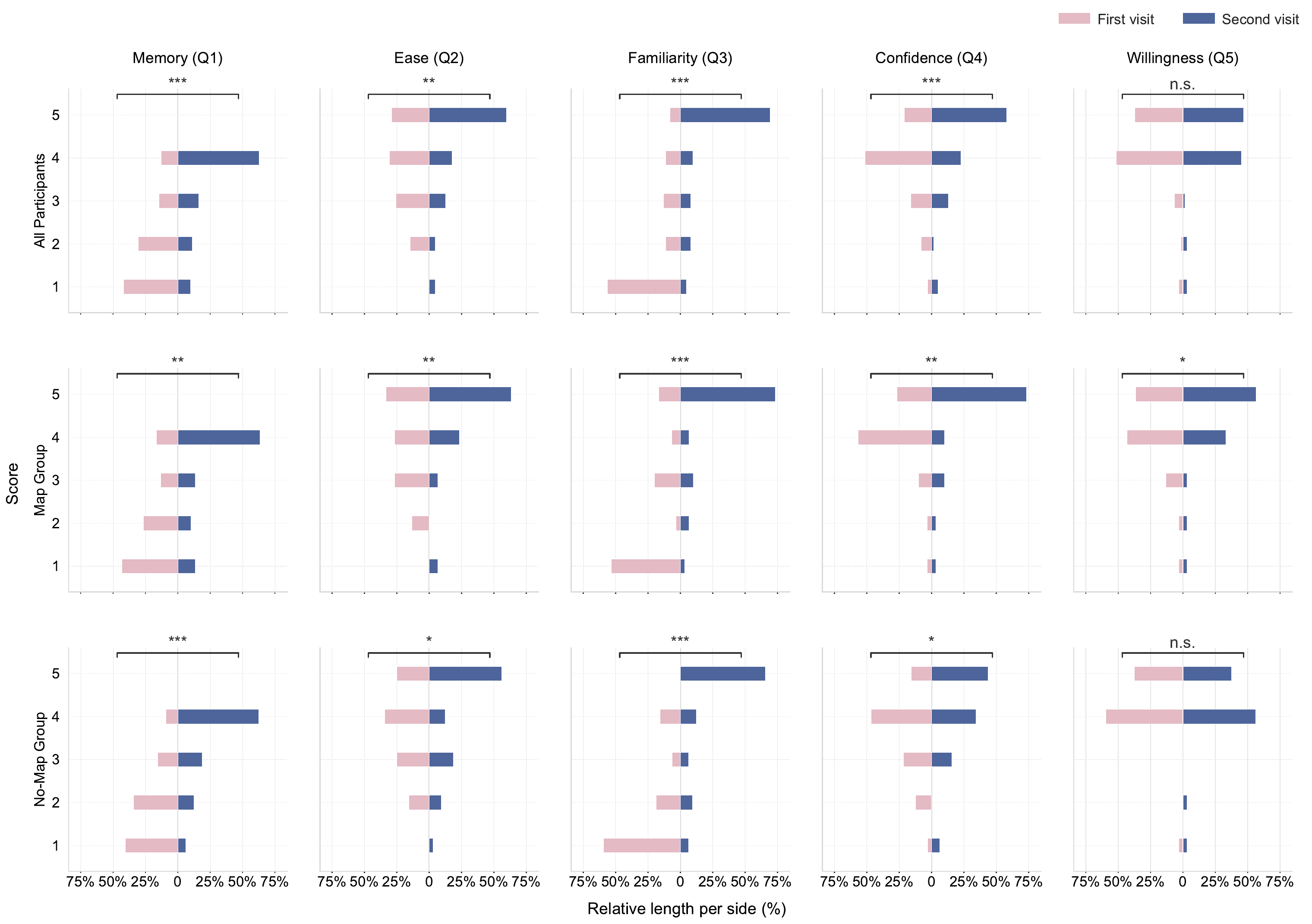} 
  \caption{
In-app survey ratings for locating the same AED during general game sessions. 
Each panel shows bidirectional half-bar plots of Likert scores: the left side represents the first visit (pink) and the right side the second visit (navy). 
Bar length within each rating indicates the percentage of responses on that side. 
Top row: all participants combined; middle row: Map group; bottom row: No-Map group. 
Significance brackets indicate between-visit contrasts: *~$p<0.05$, **~$p<0.01$, ***~$p<0.001$; n.s., not significant.
}
  \Description{Bar chart comparing first and second visit scores; a red bracket shows overall significance across all groups.}
  \label{fig:firstsecondcam}
\end{figure}

We examined within-participant changes between the first and second visits to the same AED across five in-app ratings: Memory (Q1), Ease (Q2), Familiarity (Q3), Confidence (Q4), and Willingness (Q5), each rated on a Likert scale. To avoid potential bias from participants’ prior exposure during the Pre-exam, we further excluded trips involving AED 3 on Campus I and AED 7 on Campus II from this analysis.
For each question, we tested whether participants’ second-visit ratings were significantly higher than their first-visit ratings. 
The null hypothesis ($H_0$) stated that the median within-participant change was zero, whereas the alternative ($H_a$) posited a positive change. 
We applied one-tailed paired Wilcoxon signed-rank tests, which are appropriate for ordinal paired data. We controlled the family-wise error rate across the five questions using the Holm correction, applied separately to the pooled sample and to each condition (Map and No-Map).
Figure~\ref{fig:firstsecondcam} visualizes the distributions of first- and second-visit ratings for all participants, the Map group, and the No-Map group. 
Table~\ref{tab:all_participants_apped} summarizes medians, interquartile ranges, Hodges–Lehmann (HL) estimates of change, and adjusted $p$-values.

 \begin{itemize}
  \item \textbf{Memory (Q1).}
Ratings increased substantially from the first to the second visit across all groups.
For all participants, the median rose from 2.0 (IQR 1.0–3.0) to 4.0 (3.0–4.0), with an HL change of 1.5 [1.0, 1.5], $p<0.001$.
The Map group showed a similar gain (HL = 1.5 [0.5, 2.0], $p<0.01$), and the No‑Map group exhibited a nearly identical improvement (HL = 1.5 [1.0, 2.0], $p<0.001$).
These results indicate that participants remembered the AED location more clearly on their second visit, regardless of map condition.

\item \textbf{Ease (Q2).}
Ease ratings also improved overall.
The median increased from 4.0 (3.0–5.0) to 5.0 (4.0–5.0), HL = 0.5 [0.0, 1.0], $p<0.01$.
Both Map and No‑Map groups showed parallel upward shifts (HL = 0.5 [0.0, 1.0], $p<0.01$; and HL = 0.5 [0.0, 1.0], $p<0.05$, respectively).
Participants found the second visit easier to complete, suggesting growing procedural fluency.

  \item \textbf{Familiarity (Q3).}
Familiarity exhibited the largest improvement among all questions.
For all participants, the median increased from 1.0 (1.0–3.0) to 5.0 (4.0–5.0), HL = 2.5 [2.0, 3.0], $p<0.001$.
Both Map and No‑Map groups showed strong gains (HL = 2.0 [1.5, 3.0], $p<0.001$; HL = 2.5 [2.0, 3.0], $p<0.001$).
This pronounced rise reflects a substantial increase in perceived familiarity with the AED location after repeated exposure.

  \item \textbf{Confidence (Q4).}
Confidence ratings increased moderately but significantly.
Across all participants, the median rose from 4.0 (3.0–4.0) to 5.0 (4.0–5.0), HL = 0.5 [0.5, 1.0], $p<0.001$.
The Map group showed a similar pattern (HL = 0.5 [0.0, 1.0], $p<0.01$), and the No‑Map group also improved (HL = 0.5 [0.0, 1.0], $p<0.05$).
These consistent gains suggest that repeated practice enhanced participants’ confidence in locating and accessing the AED.

  \item \textbf{Willingness (Q5).}
Willingness remained high across both visits, showing a ceiling effect.
For all participants, the median stayed at 4.0 (4.0–5.0) for both visits, with no significant change (HL = 0.0 [0.0, 0.0]).
The Map group showed a small but significant increase (HL = 0.0 [0.0, 0.5], $p<0.05$), whereas the No‑Map group remained unchanged (HL = 0.0 [0.0, 0.0], $p=0.517$).
Overall willingness to engage was already near the maximum, leaving little room for further improvement.

\end{itemize}

\subsection{Survey Outcomes Before and After the Experiment}
To complement the behavioral and cognitive analyses, this section examines participants’ subjective experiences, motivations, and perceived changes before and after the experiment, providing a holistic understanding of how AEDHunter influenced both performance and perception.

\subsubsection{Detailed Feedback during Experiment}
Our participants reported that the uniformity of building exteriors and interiors, where similarly styled lobbies, corridors, and AED cabinets reduced the distinctiveness of visual landmarks, disrupted their ability to form clear spatial memories of each device’s location. Many AEDs were installed at or near ground level and thus below typical sightlines, were frequently overlooked during timed searches, costing critical seconds as participants scanned low-lying areas. Moreover, several participants noted that a growing sense of competence increased their willingness to assist in OHCA cases.

\subsubsection{Motivation of Attending Experiment}
Participants’ motivations for joining the experiment were collected, with multiple selections allowed from the three predefined options listed in Table~\ref{table:motiv}, and all participants selected at least one option. ``For fun'' was the most frequently selected motivation, chosen by 75\% of respondents. ``For the reward'' was selected by 30\% of participants. Notably, 10\% of participants reported the reward as their sole motivation for participation, while 40\% indicated that fun was their only reason.

\begin{table}[tb] 
  \centering
  \caption{Participants’ motivations for joining the experiment with multiple selections allowed.}
  \Description{}
  \label{table:motiv}
  \begin{tabular}{lcc}
      \toprule
      \textbf{Motivation} & \textbf{Any Selection \% (n/N)} & \textbf{Sole Selection \% (n/N)} \\
      \midrule
       For fun	          &75 (15/20) &	40 (8/20)\\
       For the reward	  &30 (6/20)  &	10 (2/20)\\
       Out of curiosity  &30 ( 6/20)  &	10 (2/20)\\
      \bottomrule
  \end{tabular}
\end{table}

\subsubsection{Awareness of AED Locations Before and After the Experiment.}

\begin{figure}[tb]
  \centering
  \includegraphics[width=0.7\linewidth]{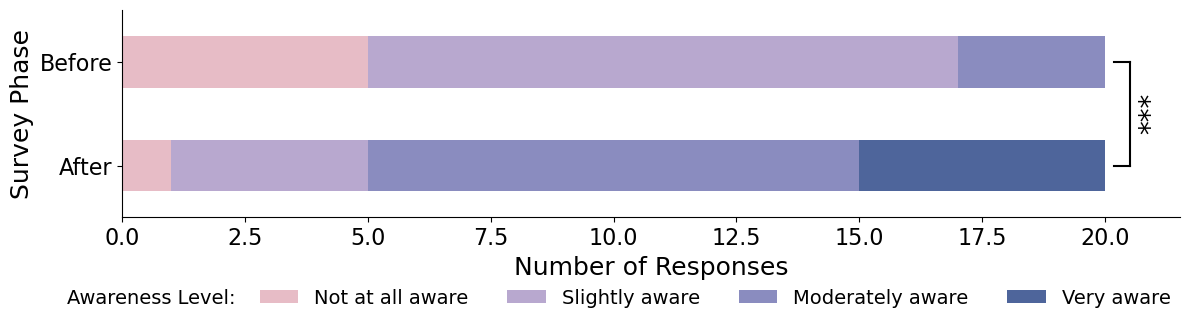} 
  \caption{ Stacked bar chart of participants' self-reported AED awareness outside the campus before and after attending the experiment. Asterisks indicate statistical significance
(*** $p < 0.001$).} 
  \Description{ Stacked bar chart of participants' self-reported AED awareness outside the campus before and after attending the experiment. Asterisks indicate statistical significance
(*** $p < 0.001$)} 
  \label{fig:awareness_compa}
\end{figure}

We assessed changes in participants’ awareness of AED locations outside the experimental areas before and after attending the study, as shown in Figure~\ref{fig:awareness_compa}. 
The data indicated a positive shift in awareness levels. 
Notably, individuals who initially reported being ``Not at all aware'' showed considerable improvement, with some moving to ``Slightly aware'' or ``Moderately aware''.
Similarly, those who began as ``Slightly aware'' often progressed to higher awareness levels. 
A particularly striking finding was the rise in the ``Very aware'' category, defined as actively looking for AEDs.
Before the intervention, no participant reported this proactive behavior, but afterward, the proportion increased substantially, suggesting that AEDHunter not only moved participants from passive noticing to moderate awareness but also fostered an intentional practice of scanning the environment and deliberately identifying AED locations.

To quantify this change, the five awareness categories were coded from 1 (``Not at all aware'') to 5 (``Extremely aware''), and a one-tailed paired Wilcoxon signed-rank test was conducted on the pre- and post-intervention scores. 
The null hypothesis ($H_0$) stated that the median within-participant difference was zero, whereas the alternative ($H_a$) posited that the median difference (post–pre) was greater than zero, indicating an increase in awareness. 
The test revealed a statistically significant increase ($p<0.001$), with median scores rising from 2.00 (IQR~1.75–2.00) before the intervention to 3.00 (IQR~2.75–3.25) afterward. 
The median change (post–pre) was 1.00 (IQR~1.00–1.25). 
These results indicate that AEDHunter was associated with a statistically significant and meaningful enhancement in participants’ awareness of AED locations beyond the experimental settings.

\subsubsection{Encouragement Ratings for AEDHunter Elements.}

To assess how different in-app components motivated rapid AED retrieval, participants rated three specific elements on a 5-point Likert scale (1 = Not at all encouraging, 5 = Extremely encouraging). 
The evaluated elements were: (i) the on-screen figure indicating that someone needs an AED, (ii) the elapsed time display, and (iii) the survival probability indicator. 
The on-screen figure received a median rating of 2 (IQR~[1.75,~4.25]), suggesting a weaker motivational effect compared with the other two cues. 
In contrast, the elapsed time display and the survival probability indicator attained higher median ratings of 4 (IQR~[4,~5]) and 5 (IQR~[4.75,~5]), respectively. 

Two-tailed paired Wilcoxon signed-rank tests with Holm correction for multiple comparisons were conducted. 
For each comparison, the null hypothesis ($H_0$) stated that the median paired difference between conditions was zero (no preference), whereas the alternative ($H_a$) posited that it was non-zero. 
The tests indicated that both the elapsed time display and the survival probability indicator were rated significantly more encouraging than the on-screen figure ($p_{\text{adj}}<0.01$), whereas the difference between the two dynamic cues was not significant ($p_{\text{adj}}=0.143$).

\subsubsection{Usability of AEDHunter}
Participants evaluated their overall experience using the System Usability Scale (SUS)~\cite{brooke1996sus}. 
The application achieved an overall SUS score of 83.1 ± 7.6 out of 100, indicating a generally positive perception of usability among participants.

\subsection{Between-group Improvement Analysis}

We compared the extent of performance improvement between the Map and No-Map groups using two-tailed Mann–Whitney U tests on (a) the change in retrieval duration between the Pre-exam and Post-exam~I sessions, and (b) the change in five in-app rating scores between the first and second visits. This nonparametric approach allowed for robust comparisons of group differences without assuming normality.

For retrieval duration, the null hypothesis ($H_0$) stated that the Map and No-Map groups shared the same distribution of pre–post change scores, whereas the alternative ($H_a$) posited that these distributions differed. For the survey ratings, separate tests were conducted for each of the five items (Q1–Q5), with $H_0$ assuming identical distributions of first–second visit changes across groups and $H_a$ indicating a difference. All comparisons were two-tailed, and the Holm correction was applied across Q1–Q5 to control for the family-wise error rate.

For retrieval-duration reduction, the two-tailed Mann–Whitney U test revealed no significant difference between the Map and No-Map groups ($p=0.820$), and the median difference in change scores (Map–No-Map) was 0.4 s. 
For the survey items, two-tailed Mann–Whitney U tests with Holm correction across Q1–Q5 likewise showed no significant between-group differences (all $p_{\text{adj}}\ge0.553$). The median differences (Map–No-Map) were Q1~=~0.00, Q2~=~+0.50, Q3~=~–1.00, Q4~=~0.00, and Q5~=~0.00, indicating minimal variation across items.

\section{DISCUSSION} 
Our findings demonstrate that AEDHunter’s game-based training significantly improved participants’ ability to locate AEDs, strengthened their spatial awareness, and enhanced their confidence in emergency response. 
Participants retrieved AEDs more efficiently after the intervention, and both behavioral logs and survey responses revealed meaningful changes in attention, strategy, and motivation. 
This discussion addresses three dimensions: 
(1) the behavioral and cognitive processes underlying these improvements, including exploratory observations of long-term retention;
(2) the design and practical implications for scalable AED awareness training; and  
(3) the study’s limitations and directions for future research.

\subsection{Behavioral and Cognitive Processes Underlying Improvement}

To understand how AEDHunter training led to performance improvements, we interpret participants’ gains through a set of complementary lenses. We first characterize how retrieval efficiency changes, and then use sensor-based phase segmentation and micro-level movement states to localize where time savings and hesitation reduction occur. We subsequently discuss training-induced shifts in attentional awareness, confidence, and willingness to intervene, and finally consider contextual moderators (environmental layout and guidance modality, e.g., Map vs.\ No-Map) as well as exploratory evidence of longer-term memory retention and decay.

\subsubsection{Behavioral Efficiency Gains}
In our experiments, the training shortened AED retrieval times for most participants and transformed participants’ search behavior. Rather than merely becoming faster, participants demonstrated more strategic and direct movement patterns, especially in the Building Search phase. Consistently, the relative pause-duration reduction ($\Delta D_P$) indicates that most participants reduced their pause time during the Post-exam~I session, suggesting that they had become more familiar with the environment. These behavioral changes suggest that participants had internalized AED locations and reduced their reliance on conscious planning, indicating stronger spatial memory and environmental familiarity. In cardiac emergencies, every second counts; each minute of defibrillation delay is known to decrease survival chances by about 7–10\%. Consequently, these efficiency gains have the potential to translate into tangible improvements in real‑world life‑saving outcomes.

\subsubsection{Sensor-based Behavioral Insights}
By examining retrieval trips from two complementary perspectives, a macro-level segmentation of each trip into Preparation (optional), Building Search, and Indoor AED Search phases, and a micro-level classification of moving versus exploratory pausing states, we gain nuanced insights into how training transforms behavior in emergency simulations. The near disappearance of the Preparation phase in the Post-exam I suggests that once participants have internalized AED locations, they devote significantly less cognitive and physical effort to planning before setting off. Meanwhile, the Building Search and Indoor AED Search phases remain necessary but become more focused and efficient, indicating increased confidence and directional certainty.

At a finer scale, the sensor-based movement classification shows that the total time spent pausing, which is defined as moments when participants slowed down or stood still, decreased modestly after training. These micro-level pauses typically indicate deliberation or uncertainty; thus, their reduction suggests that participants can navigate more fluidly and confidently following repeated practice. 
Taken together, the macro- and micro-level traces suggest reduced pre-departure planning and fewer in-route hesitations after training.

From a training design perspective, these results emphasize the value of sensor-based techniques in capturing precisely when and where participants get stuck. For example, if participants frequently pause during the Indoor AED Search phase, interactive guidance or targeted practice could address that specific challenge. Conversely, if frequent Preparation phases persist for certain trainees, additional orientation materials or structured practice might reduce pre-departure hesitations. Over time, such data-driven refinements could foster adaptive training environments where learners receive timely support the moment they need it, ultimately bolstering both speed and confidence in real-world emergency responses.

\subsubsection{Improved Attentional Awareness}

Before the intervention, most participants paid little attention to AEDs in their daily environments, reflecting a general lack of awareness of these critical safety devices. 
Following the training, many participants became more attentive to AED locations and a few participants even reported intentionally noticing them in their surroundings.
This shift from low attention to heightened attention suggests that increasing the personal relevance of AEDs and embedding them in everyday contexts can help raise awareness.
This pattern aligns with selective attention theory and prior work on personal relevance and situational salience~\cite{conway2000construction, sui2019self}, which indicate that stimuli linked to meaningful or self-relevant contexts are more likely to capture and sustain attention. 
By rendering AEDs personally salient through hands‑on activities, the intervention encouraged participants to detect and remember these devices in everyday contexts. The shift from casual perception to purposeful search therefore transformed situational awareness into a habitual safety practice, enabling participants to locate critical equipment rapidly during real‑world emergencies.

\subsubsection{Confidence and Willingness to Intervene}
Participants’ self-reported confidence in locating an AED increased significantly during the second visit to the same target, whereas their willingness to help remained consistently high. This pattern was broadly similar across both the Map and No-Map groups (Table~\ref{tab:all_participants_apped} in Appendix~\ref{A_appsurvey}). 
Because routine sessions could start at different times and locations, the second visit was not a simple replay of the first route. Practice under these conditions likely strengthened cues anchored to the goal area, such as the entrance, floor, or nearby landmarks, rather than every possible route. A modest, consistent rise in confidence is therefore plausible. The direction of change was consistent across groups, and descriptively, map support did not appear to produce a markedly different magnitude of change.

From an application perspective, two implications emerge. 
First, AEDHunter lowers the barrier to frequent, in-situ practice, allowing users to repeatedly calibrate real-world location cues and, over time, feel more certain they can find the device even when starting from different places. 
Second, willingness to intervene was already high and changed little, suggesting that the near-term bottleneck is not intention but the ability to locate the AED quickly. 
Willingness is also shaped by factors beyond navigation, such as perceived risk, responsibility, or device-use confidence, so pairing AEDHunter with cardiopulmonary resuscitation (CPR) or AED skills training and basic legal or safety information could address those aspects.

Moreover, 70\% of participants reported joining ``for fun'' or ``out of curiosity'' rather than for external rewards, suggesting that an app-based, gamified approach can tap into intrinsic motivation for learning an emergency skill. 
This enjoyment factor, combined with immediate performance feedback (e.g., survival probability), appears to strengthen participants’ affective engagement with the activity. 
Conversely, 10\% of respondents were motivated primarily by monetary or external incentives, indicating that voluntary participation could be further enhanced through well-structured rewards, particularly for harder-to-reach populations.

\subsubsection{Environmental Context and Layout Effects}
Campus~II housed its AED in a library with transparent outer walls. The library is a public facility that many participants may have visited before and already knew how to locate it. Preexisting knowledge of the building's layout, together with immediate visibility into the interior, likely contributed to the smaller yet meaningful time reductions observed on Campus~II. By contrast, Campus~I's AED was placed on a wall near the entrance of a standard research building; this environment was less visually open and less familiar, creating greater opportunity for participants to benefit from game-based training.

\subsubsection{Learning Outcomes Across Map and No-Map Groups}

The total retrieval duration decreased from Pre- to Post-exam~I in both groups, as reflected by positive relative retrieval-duration reductions ($\Delta D_T$), confirming the presence of learning. 
When decomposed by trip phase, the No-Map group showed improvement during the Building Search phase, whereas the Map group exhibited significant improvement mainly during the Preparation phase. 
The Preparation phase lasted only a few seconds at Pre-exam, so even statistically significant changes had limited impact on total trip duration. 
In contrast, Building Search was the longest segment, and improvements in this phase contributed more substantially to overall efficiency.
The median and interquartile range of Post-exam~I durations were similar across groups, suggesting that the observed effects primarily reflect general acceleration due to practice, while the interfaces differed mainly in which task stages the time savings occurred. 
Consistent with these behavioral results, self-reported ratings of memory, ease, familiarity, confidence, and willingness showed similar patterns of improvement across both groups.

\subsubsection{Memory Retention and Decay}

In the main study, participants completed three exams. Campus~I participants used AED 3 for both the Pre-exam and Post-exam~I (the practiced AED) and AED 5 for Post-exam~II (the novel AED). 
Six to seven months later, three Campus~I participants returned for two follow-up sessions (Follow-up exam~I and Follow-up exam~II) that replicated Post-exam~I and Post-exam~II. Each session began at the same designated starting point and used the same interface and AED locations as in the main study.

We conducted two complementary analyses to interpret the follow-up outcomes. 
First, for Follow-up exam~I (practiced AED), we assessed objective retention by comparing each participant’s retrieval time with their own Pre-exam and Post-exam~I baselines. 
Second, participants reflected on their subjective experiences across Follow-up exam~I and~II (perceived difficulty and confidence), providing a within-participant contrast between practiced and novel accessibility.

For the practiced AED, retrieval time improved from the Pre-exam to Post-exam~I and showed a partial rebound at Follow-up exam~I, yet remained faster than in the Pre-exam for all participants (Figure~\ref{fig:folo}a in Appendix~\ref{appendix:followupfig}). 
In interviews, participants reported that at the beginning of Follow-up exam~I, they were uncertain about the exact location of the target building; they remembered only its approximate direction and gradually recalled the precise location as they approached it. 
By contrast, participants exhibited weaker recall of the novel AED in Follow-up exam~II (Figure~\ref{fig:folo}b in Appendix~\ref{appendix:followupfig}), consistent with their limited exposure during the main study.

These results suggest that participants retained partial spatial memory of the practiced AED even after several months. 
Retrieval times provide behavioral evidence that spatial memory for the practiced AED remained accessible even after several months, although its precision may have declined over time (e.g., retaining coarse directional cues rather than exact location), whereas memory for the novel AED stayed weak, possibly due to insufficient repetition for consolidation.

\subsection{Design Implications}
Our findings indicate that gamified mobile applications can effectively engage the public and are associated with improved individual preparedness for AED retrieval. In line with recent work~\cite{hognogi2023role}, this approach leverages mobile technologies to actively involve citizens in addressing urgent societal challenges and strengthening community capabilities. 
Building on these insights, future research should explore strategies to further enhance public awareness, attention, and spatial memory of AED locations, ultimately improving emergency response readiness. Four promising directions emerge:
\begin{itemize}
\item \textbf{Distinctive Signage and Markings:} 
Eye-catching signage can significantly improve AED discoverability. Using vibrant colors, distinctive shapes, reflective stickers, or LED lighting around storage areas can help devices stand out in otherwise uniform settings, complementing digital training with environmental cues.
\item \textbf{Multi-Modal Environmental Cues:} 
Because indoor search constitutes a substantial portion of retrieval time, additional sensory cues may facilitate faster detection. For example, incorporating localized audio or haptic feedback that activates when a participant nears an AED could draw attention more effectively, particularly in visually cluttered, crowded, or unfamiliar environments.
\item \textbf{Real-Time Classifier Integration in AEDHunter:} 
Embedding a lightweight classifier (e.g., SVM) to provide continuous sensor-based inferences could enable the system to trigger micro-prompts (e.g., ``You seem stuck, try the lobby alcove'') precisely when needed.
This approach mirrors just-in-time adaptive interventions, guiding participants more efficiently toward the AED.
\item \textbf{Guidance Modality (Map vs.\ No-Map):} 
In our study, the two guidance modalities did not exhibit statistically significant differences in pre–post change scores for total retrieval time from Pre-exam to Post-exam~I.
From a deployment standpoint, this finding suggests that effective training can occur both with and without a map overlay. 
A distance/time–only mode remains practical for resource-limited scenarios and large-scale deployments, whereas map-based hints can serve as adaptive support for onboarding novices or assisting navigation in complex environments.
\end{itemize}

\subsection{Limitations and Future Work}
Our experiments were conducted with young university participants who were familiar with smartphones in campus environments. These choices limit generalizability to other populations (e.g., older adults, children, or people with limited smartphone experience) and to public settings with different building layouts, signage conventions, access constraints, and AED placement policies. In addition, the relatively small sample size may limit the robustness of our estimates. Future work should therefore include larger, multi-site deployments with more diverse participants and environments to validate the observed effects.

Our long-term assessment was limited in scope. In a follow-up conducted approximately six to seven months after the main study, the results suggest that improvements in retrieval efficiency have persisted to some extent. However, only three participants completed the follow-up, substantially limiting the strength of our conclusions. This finding highlights the need for longitudinal studies with larger cohorts to systematically characterize retention. Building on the spacing effect~\cite{roediger2011critical}, periodic challenges to locate nearby AEDs may further enhance long-term familiarity and confidence. Extending the training period beyond three weeks could also help clarify how training duration influences acquisition and retention, enabling more robust comparisons across long-term intervention designs.

Finally, our experiments were conducted in low-risk, repeatable training settings and did not reproduce the psychological or social stress of real OHCA situations. Thus, improvements should be interpreted as gains in training-context efficiency and familiarity and may be optimistic relative to field conditions. Prior work suggests an inverted-U relationship between stress and performance: moderate urgency can mobilize attention, whereas excessive stress can narrow attention and increase working-memory load~\cite{yerkes1908relation}. Future studies should evaluate how well these gains transfer to high-stress scenarios. While shorter AED retrieval times could plausibly improve response, evaluating effects on survival is beyond the scope of this work and would require dedicated clinical studies.

\section{CONCLUSION}
This paper presents AEDHunter, a location-based mobile game that transforms the life-saving yet underutilized task of locating an AED into an engaging, data-rich micro-adventure. By combining clinically grounded time pressure with real-time sensor logging (e.g., GPS, Wi-Fi, IMU, barometer, and compass) and a lightweight two-state classifier that identifies exploratory pauses, which are used as a behavioral learning signal, AEDHunter delivers two key outcomes: (i) significantly faster AED retrieval, with a median within-participant relative retrieval-duration reduction of $\Delta D_T = 0.39$ and a median relative pause-duration reduction of $\Delta D_P = 0.40$ after only a few training sessions; and (ii) increased participant confidence in retrieving AEDs, accompanied by improved awareness and proactive surveying of their environment for AED locations. These results suggest that low-cost, gamified training delivered through everyday smartphones can help reclaim life‑critical seconds and cultivate a preparedness mindset at the individual level, representing a promising avenue for improving bystander AED use that deserves further exploration.

\begin{acks}
We sincerely thank all reviewers for their insightful suggestions. This work was supported by JST PRESTO (Grant Number JPMJPR2527), JSPS KAKENHI (Grant Number JP20H00622), and JST SPRING (Grant Number JPMJSP2108).
\end{acks}

\bibliographystyle{ACM-Reference-Format}
\bibliography{sample-base}

\clearpage
\appendix

\section{Appendices}
This appendix presents supplementary results, including the performance of the exploratory pausing behavior detection models, the outcomes of the in-app survey, and the results from the long-term follow-up study.

\subsection{Classification Performance of Exploratory Pausing Behavior Detection}
\label{appendix:b}
\begin{table}[h]  
  \caption{Comparison of performance metrics across machine learning models in behavior detection.}
  \label{table:mobile_detection}
  \centering
  \begin{tabular}{llrrr}
  \toprule
  \textbf{Model} & \textbf{Label} & \textbf{Precision} & \textbf{Recall} & \textbf{F1-score} \\
  \midrule
  \multirow{2}{*}{Random Forest} 
  & Exploratory Pausing & \textbf{0.81} &0.79& 0.80 \\
  & Moving             &0.96& 0.97 & \textbf{0.97} \\
  \cmidrule(lr){1-5}
  \multirow{2}{*}{XGBoost} 
  & Exploratory Pausing & 0.77 & 0.82 & 0.79 \\
  & Moving              & \textbf{0.97} & 0.96 & 0.96 \\
  \cmidrule(lr){1-5}
  \multirow{2}{*}{Support Vector Machine} 
  & Exploratory Pausing & 0.80 & \textbf{0.84} & \textbf{0.82} \\
  & Moving              & \textbf{0.97} & 0.96 & \textbf{0.97} \\
  \bottomrule
  \end{tabular}
\end{table}

\subsection{Score of In-App Survey}\label{A_appsurvey}
\label{appendix:a}
\begin{table}[h]
\centering
\caption{In‑app survey ratings comparing participants’ first and second visits to the same AED during general game sessions. 
Values are medians with interquartile ranges (Q1–Q3). 
Median Change denotes the Hodges–Lehmann estimate of paired differences (Second – First) with 95\% bootstrap confidence intervals. 
Significance was tested using one‑tailed paired Wilcoxon signed‑rank tests with Holm correction within each group. * $p<0.05$, ** $p<0.01$, *** $p<0.001$.}
\label{tab:all_participants_apped}

\begin{tabular}{llrcccccc}
\toprule
\textbf{Group} & \textbf{Question} & 
\shortstack{\textbf{First visit}\\\textbf{Median (IQR)}} &
\shortstack{\textbf{Second visit}\\\textbf{Median (IQR)}} &
\shortstack{\textbf{HL change}\\\textbf{(Second$-$First)}} &
\shortstack{\textbf{95\% CI}\\\textbf{(HL)}} &
\textbf{Adj. }$\boldsymbol{p}$ \textbf{-value}\\
\midrule
\multirow{5}{*}{All} 
 & Q1 & 2.0 (1.0, 3.0) & 4.0 (3.0, 4.0) & 1.5 & [1.0, 1.5] & ***\\
 & Q2 &  4.0 (3.0, 5.0) & 5.0 (4.0, 5.0) & 0.5 & [0.0, 1.0] & **\\
 & Q3 &1.0 (1.0, 3.0) & 5.0 (4.0, 5.0) & 2.5 & [2.0, 3.0] & ***\\
 & Q4 & 4.0 (3.0, 4.0) & 5.0 (4.0, 5.0) & 0.5 & [0.5, 1.0] & ***\\
 & Q5 &4.0 (4.0, 5.0) & 4.0 (4.0, 5.0) & 0.0 & [0.0, 0.0] &  0.075\\
\midrule
\multirow{5}{*}{Map} 
 & Q1 &  2.0 (1.0, 3.0) & 4.0 (3.0, 4.0) & 1.5 & [0.5, 2.0]&  **\\
 & Q2 &  4.0 (3.0, 5.0) & 5.0 (4.0, 5.0) & 0.5 & [0.0, 1.0] & **\\
 & Q3 & 1.0 (1.0, 3.0) & 5.0 (4.2, 5.0) & 2.0 & [1.5, 3.0] & ***\\
 & Q4 & 4.0 (4.0, 4.8) & 5.0 (4.2, 5.0) & 0.5 & [0.0, 1.0] & **\\
 & Q5 &  4.0 (4.0, 5.0) & 5.0 (4.0, 5.0) & 0.0 & [0.0, 0.5] & *\\
\midrule
\multirow{5}{*}{No-Map} 
 & Q1 & 2.0 (1.0, 2.2) & 4.0 (3.0, 4.0) & 1.5 & [1.0, 2.0] & ***\\
 & Q2 & 4.0 (3.0, 4.2) & 5.0 (3.0, 5.0) & 0.5 & [0.0, 1.0]  & *\\
 & Q3 & 1.0 (1.0, 2.0) & 5.0 (4.0, 5.0) & 2.5 & [2.0, 3.0] & ***\\
 & Q4 & 4.0 (3.0, 4.0) & 4.0 (4.0, 5.0) & 0.5 & [0.0, 1.0] & *\\
 & Q5 & 4.0 (4.0, 5.0) & 4.0 (4.0, 5.0) & 0.0 & [0.0, 0.0]  & 0.517\\
\bottomrule
\end{tabular}
\end{table}

\subsection{Supplementary Results of Long-Term Follow-Up}

\label{appendix:followupfig}
\begin{figure}[hb]
  \centering
  \includegraphics[width=0.92\linewidth]{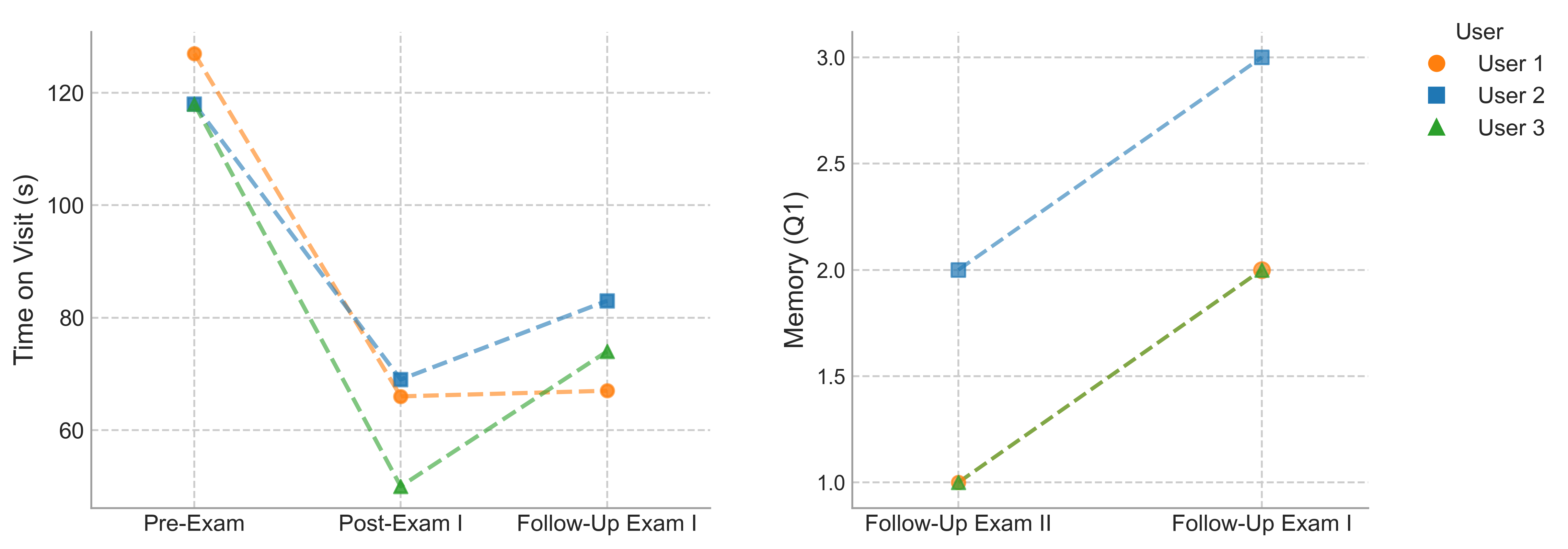}  
  \caption{
Results for participants with follow-up data. (a) illustrates retrieval time for the practiced AED across the Pre-exam, Post-exam I, and Follow-up exam I among participants who completed follow-up exams. (b) depicts Memory (Q1) scores at Follow-up exam I (for the practiced AED) and Follow-up exam II (for the novel AED) for the same participants. Lines connect within-participant measurements to highlight changes over time.}
  \Description{Results of follow-up experiments}
  \label{fig:folo}
\end{figure}

\end{document}